\documentclass[aps,journal=prl,twocolumn,noeprint,superscriptaddress,amsmath,amssymb,showpacs,nolongbibliography]{revtex4-1}
\usepackage{graphicx}
\usepackage{amsmath}
\usepackage{graphics}

\usepackage{amsmath}
\usepackage{amsfonts}
\usepackage{amssymb}
\usepackage{xcolor}
\usepackage{float}
\graphicspath{{figs/}}

\usepackage[caption=false]{subfig}
\newcommand{\phantomsubfloat}[1]{
    {
        \captionsetup[subfigure]{labelformat=empty}
        \subfloat[][]{#1}
    }%
}

\usepackage[bookmarks=false]{hyperref}
\hypersetup{colorlinks=true, citecolor=blue, urlcolor=blue, linkcolor=blue}

\newcommand{\pd}{\phantom{\dagger}}
\newcommand{\figref}[1]{\ref{#1}}

\usepackage[normalem]{ulem}

\begin{document}

\title{A General Non-Markovian Electronic Friction and Langevin Dynamics Framework for Nonequilibrium Systems}

\author{S.\ L.\ Rudge}
\email{Email: samuel.rudge@physik.uni-freiburg.de}
\affiliation{Institute of Physics, University of Freiburg, Hermann-Herder-Str. 3, D-79104 Freiburg, Germany}
\author{R.\ J.\ Preston}
\affiliation{Institute of Physics, University of Freiburg, Hermann-Herder-Str. 3, D-79104 Freiburg, Germany}
\author{D.\ S.\ Kosov}
\affiliation{College of Science and Engineering, James Cook University, Townsville, QLD, 4811, Australia }
\author{M.\ Thoss}
\affiliation{Institute of Physics, University of Freiburg, Hermann-Herder-Str. 3, D-79104 Freiburg, Germany}

\begin{abstract}

\noindent We present a general framework for simulating the nonequilibrium vibronic dynamics of molecules interacting with metal surfaces, which utilizes non-Markovian electronic friction and the corresponding generalized Langevin equation. The method employs a Markovian embedding scheme to sample coordinate-dependent quantum colored noise, with the underlying memory kernels represented systematically by exponential decompositions obtained directly from the electronic forces. In doing so, the approach extends the widely used electronic friction and Langevin dynamics formalism beyond the conventional Markovian approximation while retaining the coordinate dependence of the electronic back-action. We demonstrate the method for nonequilibrium charge transport through a vibrationally coupled molecular junction, where comparisons with numerically exact hierarchical equations of motion simulations show that the non-Markovian formulation is both more accurate and more robust than its Markovian counterpart. In particular, we find that non-Markovian electronic forces play a decisive role in the nonequilibrium vibrational dynamics, stabilizing regimes in which the conventional Markovian approximation fails.
\end{abstract}

\maketitle

\newpage

\section{Introduction} \label{sec: Introduction}

\noindent The vibronic dynamics of molecules interacting with metal surfaces is crucial to understanding a wide variety of scenarios in chemistry and physics. These include scattering of molecules off metals \cite{Preston2025_nonadiabatic,Buenermann2015,Maurer2019_hot}, reaction and desorption dynamics at surfaces \cite{Tully2000_chemical,Erpenbeck2019,Saalfrank2006,Brandbyge1995}, and nonequilibrium transport through molecular nanojunctions \cite{Ke2023_current,Erpenbeck2018,Preston2023,Erpenbeck2023_how}. However, simulating such systems is generally challenging, as they often operate in strongly coupled regimes, where the dynamics are highly non-Markovian and non-perturbative. Furthermore, the continuum of electronic states in the metal allow electronic excitations of arbitrarily low energy, which naturally requires treatment beyond the Born-Oppenheimer approximation. 

Among the many approaches to simulating vibronic dynamics at metal-molecule interfaces are fully quantum methods, such as the hierarchical equations of motion (HEOM) \cite{Ke2021,Schinabeck2016,Erpenbeck2019,Tanimura2020,Jin2007,Jin2008} and the multilayer, multi-configurational time-dependent Hartree (ML-MCTDH) approaches \cite{Wang2003,Wang2009,Manthe2008,Vendrell2011,Wang2011,Wang2016}. However, while these approaches can provide useful  benchmarks for relatively small models, they are numerically challenging to use, especially in systems with many vibrational modes or large vibrational excitation \cite{Schinabeck2018,Kaspar2021}. Consequently, another class of methods has been developed, in which the molecular vibrational degrees of freedom are treated classically while under the influence of the quantum mechanical electronic degrees of freedom in the metal and the molecule.  

Examples of these mixed-quantum classical approaches are methods such as the mean-field Ehrenfest approach \cite{Bellonzi2016,Subotnik2010,Verdozzi2006,Kartsev2014,Cunningham2015,Stock2005,Dundas2009,Erpenbeck2018,Smorka2025}, surface-hopping approaches \cite{Tully1971,Tully1990,Dou2015a,Dou2015b,Dou2015c,Shenvi2008,Shenvi2009a,Shenvi2009b}, and electronic friction and Langevin dynamics (EFLD) \cite{HeadGordon1995,Maurer2016,Dou2017a,Lue2011,Lue2019,Bode2011,Bode2012}, which is the method investigated in this work. Generally, the family of EFLD approaches is obtained by assuming a further limit of weak nonadiabaticity beyond the assumption of classical molecular vibrations \cite{Chen2019a,Dou2017c}. The EFLD approach used in this work enforces this limit via a timescale separation between fast electronic relaxation and slow vibrational dynamics, yielding a Gaussian electronic influence and a non-Markovian stochastic Langevin equation in which the quantum electronic degrees of freedom appear as effective electronic forces \cite{Chen2019a,Preston2026_negative,Dou2017c}. 

In most practical applications, however, the non-Markovian Gaussian colored noise is too challenging to sample directly, necessitating further approximations. Even in the quasi-stationary limit, where the two-time electronic forces become functions only of the time difference, colored-noise sampling has only been achieved for correlation functions that either factorize into coordinate- and time-dependent components \cite{Trenins_nonmarkovian_2025} or are independent of the vibrational coordinates altogether \cite{Lawrence2019_on}. Consequently, the vast majority of studies employ the Markovian limit, in which the electronic noise reduces to white noise and can be sampled as an independent Wiener process at each time step \cite{HeadGordon1995,Maurer2016,Rudge2023,Rudge2024,Dou2015d,Dou2016a,Maeck2024}. Although this approximation has proven successful for describing nonequilibrium steady-state dynamics in simple molecular nanojunctions \cite{Rudge2024,Maeck_Vibrational2025} and scattering in the limit of heavy nuclei and strong molecule-metal coupling \cite{Preston2025_nonadiabatic,Lu2026_ahaldane}, growing evidence indicates that non-Markovian effects become essential in systems with many-body interactions \cite{Rudge2023,Rudge2024,Chen2019a,Chen2019b,Maeck2024} or bias-driven vibrationally coupled inelastic electronic transitions \cite{Preston2026_negative}.

Consequently, in this work we present a general method for propagating the non-Markovian Langevin equation without factorizing the coordinate- and time-dependence of the correlation function of the stochastic force. The approach is based on Markovian embedding techniques \cite{Lu2026_memory,Trenins_nonmarkovian_2025,Ceriotti2010_colored,Ceriotti2010_efficient,Ceriotti2012_efficient}, in which non-Markovian colored noise is represented by a set of coupled Markovian stochastic processes. In contrast to existing embedding schemes, however, our formulation retains the full coordinate dependence of the electronic forces, treats quantum rather than classical colored noise, and remains applicable out of equilibrium where the quantum fluctuation-dissipation theorem (FDT) is no longer satisfied. To achieve this, we perform independent pole decompositions of the electronic friction and stochastic-force correlation function, allowing both quantities to be represented by deterministic and stochastic auxiliary variables that are propagated alongside the generalized Langevin equation. 

To demonstrate the method, we apply it to nonequilibrium charge transport through a reduced model of a vibrationally coupled donor-acceptor nanojunction \cite{simine_vibrational_2012,segal_heat_2005,segal_heat_2006,segal_nonequilibrium_2011,Foti2018_origin,foti_interface_2017}. This system provides an ideal testbed for non-Markovian EFLD (NM-EFLD), as Ref.~\cite{Preston2026_negative} recently identified regimes of negative electronic friction associated with strong non-Markovian contributions to the electronic forces. In particular, the conventional Markovian EFLD (M-EFLD) approach was shown to become unstable in certain nonequilibrium parameter regimes. Here, we demonstrate that NM-EFLD remains stable across these regimes and accurately predicts the nonequilibrium steady-state vibrational excitation, in close agreement with numerically exact quantum benchmarks. Beyond this model, we anticipate that the proposed framework will enable the application of NM-EFLD to a broad range of molecule-metal dynamical problems, facilitated by the detailed decomposition and propagation algorithms presented here.

The paper is structured as follows. In Sec.~\ref{sec: Model}, we introduce the general molecule-metal Hamiltonian as well as the specific donor-acceptor model used as a demonstration. In Sec.~\ref{sec: Non-Markovian Langevin Dynamics}, we summarize the non-Markovian EFLD equations of motion in the quasi-stationary limit, before introducing the NM-EFLD framework in Sec.~\ref{subsec: Coordinate-Dependent Markovian Embedding} and a detailed description of the electronic force decomposition and Langevin propagation algorithms in Sec.~\ref{sec: NM-EFLD Algorithm}. In Sec.~\ref{sec: Results}, we first analyze the coordinate-dependent decomposition of the electronic forces in Sec.~\ref{subsec: Coordinate-Dependent Pole Decomposition} and then investigate nonequilibrium steady-state observables in Sec.~\ref{subsec: Transient and Steady-State Observables}. Throughout this work, we use units where $e = \hbar = 1$.

\section{Model} \label{sec: Model}

\begin{figure}
    \begin{center}
       \includegraphics[width=\columnwidth]{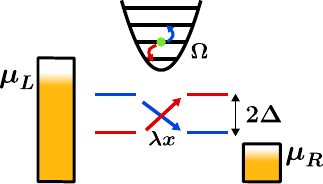}
       \caption{Schematic of the vibrationally coupled donor-acceptor model. Two electronic energy configurations are shown: $\Delta > 0$ and $\Delta <0$, which correspond to driving and dissipation of the harmonic vibrational mode, respectively.}
       \label{fig: vib energy schematic}
    \end{center}
\end{figure}

The electronic friction and Langevin dynamics approach can be applied to a wide variety of systems describing the interaction of molecules with metal surfaces. The general Hamiltonian is given by 
\begin{align}
H = \: & H_{\text{mol}} + H_{\text{met}} + H_{\text{mol-met}}, \label{eq: total hamiltonian}
\end{align}
where $H_{\text{mol}}$ is the Hamiltonian of the molecule, $H_{\text{met}}$ is the Hamiltonian of the metal surface(s), and $H_{\text{mol-met}}$ is the interaction between the two. The molecular Hamiltonian can be generally written as 
\begin{align}
    H_{\text{mol}} = \: & \sum_{m} \varepsilon^{\pd}_{m}(\hat{\boldsymbol{x}}) d^{\dag}_{m} d^{\pd}_{m} + \sum_{\substack{m \\ n\neq m}} t^{\pd}_{mn}(\hat{\boldsymbol{x}}) d^{\dag}_{m} d^{\pd}_{n} \nonumber \\
    & + \sum_{\substack{m \\ n\neq m}} U_{mn}(\hat{\boldsymbol{x}}) d^{\dag}_{m} d^{\pd}_{m} d^{\dag}_{n} d^{\pd}_{n}+ \sum_{i} \frac{\hat{p}_{i}^{2}}{2m_{i}} + U_{\text{vib}}(\hat{\boldsymbol{x}}). \label{eq: general molecular Hamiltonian}
\end{align}

Here, the electronic energies $\varepsilon^{\pd}_{m}(\hat{\boldsymbol{x}})$ represent diabatic states relevant to the dynamics, and are populated and depopulated via the creation and annihilation operators $d^{\dag}_{m}$ and $d^{\pd}_{m}$, respectively. Hopping between these states is governed by $t^{\pd}_{mn}(\hat{\boldsymbol{x}})$. Likewise, the intra- and inter-site electron-electron interaction is governed by $U_{mn}(\hat{\boldsymbol{x}})$. The electronic degrees of freedom are coupled to the vibrational modes, $\{\hat{\boldsymbol{x}},\hat{\boldsymbol{p}}\}$, which describe atomic nuclear motion in the form of normal modes, chemical bonds, or reaction coordinates. The vibrational kinetic energy is given by the second-to-last term in Eq.\eqref{eq: general molecular Hamiltonian}, while $U_{\text{vib}}(\hat{\boldsymbol{x}})$ denotes the unperturbed vibrational potential energy surface. Note that we have retained the '$\hat{}$' notation on the quantum vibrational operators to later distinguish between the mixed quantum-classical treatments and the fully quantum benchmarks. 

Although the theory presented in this work can be applied to arbitrary Hamiltonians of the form of Eq.\eqref{eq: general molecular Hamiltonian}, we will demonstrate the NM-EFLD approach specifically via the vibrationally coupled donor-acceptor model:
\begin{align}
H_{\text{mol}} = \: & \Delta \left(d^{\dag}_{1}d^{\pd}_{1} - d^{\dag}_{2}d^{\pd}_{2}\right) + \lambda \sqrt{2 m \Omega} \left(d^{\dag}_{1}d^{\pd}_{2} + d^{\dag}_{2}d^{\pd}_{1}\right)\hat{x}  
\nonumber \\
& + \frac{1}{2} m \Omega^{2}\hat{x}^{2} + \frac{\hat{p}^{2}}{2m}. \label{eq: molecular Hamiltonian donor-acceptor}
\end{align}
This model contains two electronic sites with energetic separation $2\Delta$ that are linearly coupled to a single harmonic vibration with frequency $\Omega$ and mass $m$ via the hopping term, with coupling strength $\lambda \sqrt{2m\Omega}$. 

In order to investigate the nonequilibrium behavior of the model given in Eq.\eqref{eq: molecular Hamiltonian donor-acceptor}, the molecule is coupled to two metallic surfaces or leads, which are modeled as reservoirs of noninteracting electrons with Hamiltonian 
\begin{align}
H_{\text{met}} = \: & \sum_{\alpha \in \{L,R\} }\sum_{k} \varepsilon^{\pd}_{k} c^{\dag}_{k\alpha}c^{\pd}_{k\alpha},
\end{align}
although the theory is valid for any number of surfaces. Here, the operators $c^{\dag}_{k\alpha}$ and $c^{\pd}_{k\alpha}$ create and annihilate an electron with energy $\varepsilon_{k}$ in lead $\alpha$, respectively. The surfaces are assumed to be held at local equilibrium with well-defined temperature $T$ and chemical potentials $\mu_{\alpha}$, while nonequilibrium transport conditions are generated via a symmetrically applied voltage bias across the junction, $\Phi = (\mu_{L} - \mu_{R})/e$ with $\mu_{L} = e\Phi/2 = -\mu_{R}$. 

Finally, the molecule-metal interaction term has the general form
\begin{align}
H_{\text{mol-met}} = \: & \sum_{\alpha,k} \left(V^{\pd}_{k\alpha,m}(\hat{\boldsymbol{x}})c^{\dag}_{k\alpha}d^{\pd}_{m} + V^{*}_{k\alpha,m}(\hat{\boldsymbol{x}})d^{\dag}_{m}c^{\pd}_{k\alpha}\right), \label{eq: mol-met hamiltonian}
\end{align}
where the coupling strength between state $k$ in lead $\alpha$ and state $m$ in the molecule is denoted by $V^{\pd}_{k\alpha,m}(\hat{\boldsymbol{x}})$. Although this can in principle depend on the vibrational coordinates, in the donor-acceptor model considered in this work, we explicitly assume a constant coupling strength: $V^{\pd}_{k\alpha,m}(\hat{\boldsymbol{x}}) = V^{\pd}_{k\alpha,m}$. 

The molecule-metal interaction can be further characterized by the spectral density of surface $\alpha$,
\begin{align}
    \Gamma_{\alpha,mm'}(E) = \: & 2\pi\sum_{k} V^{\pd}_{k\alpha,m}V^{*}_{k\alpha,m'} \delta(E - \varepsilon_{k}),
\end{align}
which in this work is taken to be a Lorentzian:
\begin{align}
    \Gamma_{\alpha,mm'}(E) = \: & V^{\pd}_{\alpha,m}V^{*}_{\alpha,m} \frac{W_{\alpha}^{2}}{(E - \mu_{\alpha})^{2} + W_{\alpha}^{2}}. \label{eq: Lorentzian spectral density}
\end{align}
Here, $\Gamma_{\alpha,mm'}(E)$ has a single peak centered around the chemical potential, $\mu_{\alpha}$, and bandwidth $W_{\alpha} = 10\text{eV}$. The newly defined quantities $V^{\pd}_{\alpha,m}$ now represent a constant coupling strength between lead $\alpha$ and state $m$, and they form the molecule-metal coupling strength, \\ $\Gamma_{\alpha,mm'} = 2\pi V^{\pd}_{\alpha,m}V^{*}_{\alpha,m'} $. We assume that the molecule is arranged as a chain, such that electronic site $1$ couples only to the left lead and electronic site $2$ couples only to the right lead, resulting in a spectral function diagonal in the electronic states:
\begin{align}
\Gamma_{\alpha,12} & = \: \Gamma_{\alpha,21} = \Gamma_{R,11} = \Gamma_{L,22} = 0.
\end{align}
The remaining molecule-metal couplings are set to the same value, $\Gamma_{R,22} = \Gamma_{L,11} = \Gamma_{L} = \Gamma_{R}$, with total molecule-metal coupling strength now defined as \\$\Gamma = \Gamma_{L} = \Gamma_{R}$.

A schematic of the vibrationally coupled donor-acceptor model is shown in Fig.~\figref{fig: vib energy schematic}. Two possible choices of the electronic energy configuration are also shown. At positive bias voltage and when $\Delta > 0$ (blue pathway), electrons transport inelastically through the molecule and constantly pump $2\Delta$ of energy into the harmonic vibrational mode. In contrast, electrons transporting through the opposite electronic configuration, when $\Delta < 0$ (red pathway), must dissipate energy from the vibrational mode. 

\section{Non-Markovian Electronic Friction and Langevin Dynamics} \label{sec: Non-Markovian Langevin Dynamics}

In this section, we first introduce the non-Markovian electronic friction and Langevin dynamics approach and its underlying assumptions in Sec.~\ref{subsec: Mixed Quantum-Classical Theory}. Since the method is well known in the literature, we present here only a brief outline and direct the reader to Refs.~\cite{Chen2019a,Dou2017a,Preston2026_negative} for more detailed overviews. Then, in Sec.~\ref{subsec: Coordinate-Dependent Markovian Embedding}, we discuss the Markovian embedding technique that we propose to propagate the resulting equations of motion. 

\subsection{Mixed Quantum-Classical Theory} \label{subsec: Mixed Quantum-Classical Theory}

The EFLD equations of motion can be rigorously derived by starting from a fully quantum picture of the molecular vibrations. This motivates a different decomposition of the Hamiltonian given in Eq.\eqref{eq: total hamiltonian}, in which the total system is split into vibrational and electronic components rather than molecule and metal:
\begin{align}
    H = \: & H_{\text{mol,vib}} + H_{\text{el}}(\hat{x}) \\
    H_{\text{mol-vib}} = \: & \frac{\hat{p}^{2}}{2m} + U_{\text{vib}}(\hat{x}) \\
    H_{\text{el}}(\hat{x}) = \: & H_{\text{mol-el}}(\hat{x}) + H_{\text{met}} + H_{\text{mol-met}}(\hat{x}),
\end{align}
where $H_{\text{mol-el}}(\hat{x})$ contains both the purely electronic components as well as the electronic-vibrational coupling, and the coordinate-dependence of the molecule-metal interaction has been included for the sake of generality. Note that, for the sake of simplicity, we will explicitly treat a single vibrational mode from this point onwards, although the theory can easily be extended to multiple vibrational modes.

The reduced density matrix of the molecular vibrations is obtained by tracing out the electronic degrees of freedom in the molecule and metal from the total density matrix,
\begin{align}
    \rho_{\text{vib}}(t) = \: & \text{Tr}_{\text{el}} \left\{\rho(t)\right\},
\end{align}
which can be written exactly using the Feynman-Vernon influence functional, in which the electronic influence appears as an effective action term \cite{Chen2019a,Preston2026_negative}. By expanding the influence functional to second-order in quantum deviations from the classical path  a semiclassical stochastic equation of motion for $x(t)$ is obtained,
\begin{align}
    \dot{p}(t) = \: & -\partial_{x} U_{\text{vib}} + \langle F_{\text{el}}\rangle[x(t)] + f(t),
\end{align}
In this limit, the electronic influence is described by the path-dependent mean electronic force,
\begin{align}
    \langle F_{\text{el}} \rangle [x(t)] = \: & \text{Tr}_{\text{el}}\left\{F_{\text{el}}(x(t)) \rho_{\text{el}}[x(t)]\right\},
\end{align}
as well as a Gaussian stochastic force with the colored-noise two-time correlation function,
\begin{align}
    D(t,t') = \: & \langle f(t)f(t') \rangle \nonumber \\
    = \: & \text{Tr}_{\text{el}} \left\{\left[F^{I}_{\text{el}}(x(t)),F^{I}_{\text{el}}(x(t'))\right]_{+}\rho_{\text{el}}(0)\right\}.
\end{align}
These definitions have introduced the electronic force operator
\begin{align}
    F_{\text{el}}(x(t)) = -\partial_{x} H_\text{el}(x(t)),
\end{align}
as well as the interaction picture with respect to the electronic degrees of freedom,
\begin{align}
    F^{I}_{\text{el}}(x(t)) = U^{\dag}_{\text{el}}(t,0)F_{\text{el}}(x(t))U^{\pd}_{\text{el}}(t,0),
\end{align}
where the time-ordered evolution operator is given by 
\begin{align}
    U^{\pd}_{\text{el}}(t,0) = \: & \mathcal{T} \exp \left[ -i \int^{t}_{0} d\tau  H_{\text{el}}(x(\tau)) \right].
\end{align}
and the electronic density matrix at time $t$ is a functional of the classical trajectory,  
\begin{align}
    \rho_{\text{el}}[x(t)] = \: & U^{\pd}_{\text{el}}(t,0)\rho_{\text{el}}(0)U^{\dag}_{\text{el}}(t,0).
\end{align}

Electronic friction naturally arises from $\langle F_{\text{el}} \rangle [x(t)]$ by taking a further near-adiabatic approximation \cite{Chen2019a,Preston2026_negative}, which amounts to a timescale separation between the fast electronic relaxation timescale, $\tau_{\text{el}}$, and slow vibrational time evolution, 
\begin{align}
    \langle F_{\text{el}} \rangle (t) \approx \: & F^{\text{ad}}_{\text{el}}(x(t)) - \int^{t}_{0} d\tau \: \gamma(t,\tau) \dot{x}(\tau).
\end{align}
The full non-Markovian mean electronic force has been expanded in terms of the zeroth-order or adiabatic contribution,
\begin{align}
    F^{\text{ad}}_{\text{el}}(x(t)) = \: & \text{Tr}_{\text{el}} \left\{F_{\text{el}}(x(t)) \rho^{\text{ss}}_{\text{el}}(x(t))\right\},
\end{align}
and the electronic friction, a first-order nonadiabatic correction,
\begin{align}
    \gamma(t,\tau) = \: & \theta(t - \tau) \text{Tr}_{\text{el}} \left\{F_{\text{el}}(t) U^{\pd}_{\text{el}}(t,\tau) \partial_{x} \rho^{\text{ss}}_{\text{el}}(t) U^{\dagger}_{\text{el}}(t,\tau)\right\}. \label{eq: friction exact}
\end{align}

The adiabatic contribution to the electronic force is calculated at a fixed vibrational frame $x(t)$ for $\dot{x}(t) = 0$, with the electronic degrees of freedom at the corresponding stationary state, $\rho^{\text{ss}}_{\text{el}}(x(t))$, where $\mathcal{L}_{\text{el}}(x) \rho^{\text{ss}}_{\text{el}}(x) = 0$ and $\mathcal{L}_{\text{el}}(x)\rho^{\text{ss}}_{\text{el}}(x) = -i\left[H_{\text{el}}(x),\rho^{\text{ss}}_{\text{el}}(x)\right]$. Note that, since $H_{\text{el}}(x)$ contains a dissipative fermionic bath, $\mathcal{L}_{\text{el}}(x)$ will in general admit a stationary state. In contrast to the adiabatic force, the electronic friction kernel incorporates dissipation of vibrational energy due to electron-hole pair (EHP) formation in the leads, where the $\theta(t - \tau)$ component ensures causality.

Even in this limit, the non-Markovian stochastic force is still described by a two-time correlation function, $D(t,\tau)$, and the corresponding noise is challenging to sample directly. Consequently, a further quasi-stationary approximation is often employed \cite{Preston2026_negative,Dou2017c}, assuming that the vibrations do not move far from $x(t)$ within the timescale of electronic influence, such that the electrons can be treated in the instantaneous stationary state at $x(t)$ and their dynamics depends only on relative time $t - \tau$, $U^{\pd}_{\text{el}}(t,\tau) \rightarrow e^{-iH_{\text{el}}(x(t))(t - \tau)}$, resulting in 
\begin{align}
    \gamma(x(t), t - \tau) = \: & \theta(t - \tau) \nonumber \\
    & \times \text{Tr}_{\text{el}} \left\{F_{\text{el}}(t) e^{\mathcal{L}_{\text{el}}(x(t))(t - \tau)} \partial_{x} \rho^{\text{ss}}_{\text{el}}(x(t)) \right\}. \label{eq: friction quasi-stationary} \\
    D(x(t), t - \tau) = \: & \text{Tr}_{\text{el}} \left\{F_{\text{el}}(t) e^{\mathcal{L}_{\text{el}}(x(t))(t - \tau)} \right. \nonumber \\ 
    & \qquad \left. \times \left[\delta F_{\text{el}}(t),\rho^{\text{ss}}_{\text{el}}(x(t))\right]_{+}\right\}, \label{eq: corrfunc quasi-stationary}
\end{align}
where $\delta F_{\text{el}}(t) = F_{\text{el}}(t) - F^{\text{ad}}_{\text{el}}(t)$. 

In this near-adiabatic, quasi-stationary limit, the classical vibrations now evolve according to a generalized Langevin equation,
\begin{align}
    \dot{p} = \: & - \partial_{x} U_{\text{vib}} + F^{\text{ad}}_{\text{el}}(x) \nonumber \\
    & - \frac{1}{m}\int^{t}_{0} d\tau \: \gamma(x(t) , \tau) p(t - \tau) +  f(t). \label{eq: Generalized Langevin equation}
\end{align}
which is the central equation of motion under investigation in this work. From here on, we will refer to Eq.\eqref{eq: Generalized Langevin equation} as the non-Markovian EFLD approach (NM-EFLD). 

This is contrasted with the Markovian EFLD approach (M-EFLD), which is obtained under a further Markovian approximation \cite{Chen2019a}, assuming that the electronic friction decays quickly on a timescale $\tau_{\text{el}}$ much shorter than the vibrational time-evolution, such that within $\tau_{\text{el}}$ the time evolution of $\dot{x}(t)$ is dominated by some characteristic vibrational frequency, $\dot{x}(t - \tau) \approx \text{Re}\left\{\dot{x}(t)e^{-i\Omega\tau}\right\}$, yielding 
\begin{align}
    \dot{p} = \: & - \partial_{x} U_{\text{vib}} + F^{\text{ad}}_{\text{el}}(x) \nonumber \\  
    & -\frac{1}{m}\underbrace{\int^{\infty}_{0} d\tau \: \text{Re}\left\{\gamma(x(t) , \tau)e^{-i\Omega \tau}\right\}}_{= \text{Re}\left\{\tilde{\gamma}(x,\Omega)\right\}}p(t)  +  f(t). \label{eq: Markovian Langevin equation 1}
\end{align}
Here, we can identify $\tilde{\gamma}(x,\Omega)$ as the $\Omega$-frequency component of the electronic friction spectrum,
\begin{align}
 \tilde \gamma (x,\omega) = \int^{\infty}_{-\infty} dt \: e^{-i\omega t} \gamma(x,t).
\end{align}
Note that the limits of $\tau \rightarrow \pm \infty$ naturally arise in the Markovian limit and from the causality of the electronic friction kernel. 

Although Eq.\eqref{eq: Markovian Langevin equation 1} refers to a general Markovian Langevin equation, it depends on the choice of the characteristic frequency, $\Omega$. In the majority of investigations \cite{Lue2012,Dou2017a,Rudge2024,Dou2017c}, the Markovian limit in terms of electronic friction refers specifically to the $\Omega = 0$ choice, resulting in 
\begin{align}
    \dot{p} = \: & - \partial_{x} U_{\text{vib}} + F^{\text{ad}}_{\text{el}}(x) -  \frac{1}{m}\tilde{\gamma}(x,0) p +  f(t), \label{eq: Markovian Langevin equation 2}
\end{align}
where the stochastic force is now characterized by coordinate-dependent white noise:
\begin{align}
    D(t,\tau) = \: & \tilde{D}(x,0) \delta(t - \tau).
\end{align}

In equilibrium, which in this work corresponds to zero bias voltage, the finite-frequency electronic friction and correlation function of the stochastic force satisfy the quantum fluctuation-dissipation theorem (FDT) \cite{Chen2019a},
\begin{align}
    \tilde{D}(x,\omega) = \: \omega \coth\left(\frac{\omega}{2 k_{B}T }\right)\text{Re}\left\{\tilde{\gamma}(x,\omega)\right\},
\end{align}
which reduces to the classical FDT in the Markovian limit
\begin{align}
    \tilde{D}(x,0) = \: & 2 k_{B}T \tilde{\gamma}(x,0) \label{eq: classical fdt}.
\end{align}
These properties are not guaranteed out of equilibrium, at finite bias voltage, which leads to phenomena such as current-induced heating. Specifically, at coordinates where $\tilde{D}(x,0) > 2k_{B}T\tilde{\gamma}(x,0) \geq 0$, the strength of the nonequilibrium electronic force fluctuations exceeds that of the dissipative electronic friction, resulting in net energy transfer from the electronic degrees of freedom to the vibrational mode and stochastic heating \cite{Dou2018a,Rudge2024}. 

Moreover, the positivity of the electronic friction tensor is not guaranteed at finite bias voltage, which can lead to more exotic forms of current-induced heating. For a single vibrational degree of freedom, this appears in the Markovian limit simply as negative electronic friction, $\tilde{\gamma}(x,0) < 0$, which can occur due to biased inelastic electronic transitions \cite{Preston2026_negative,Preston2022,Preston2020,Lue2011,Bode2011,Bode2012} or many-body interactions \cite{Maeck2024,Rudge2023,Rudge2024}. 

In the non-Markovian regime, the dissipative or driving nature of the electronic friction force is instead determined by the net energy exchanged over a vibrational cycle. This can be characterized by the instantaneous power dissipated from the vibrational mode by the electronic friction,
\begin{align}
    P_{\text{diss.}} = \: & \frac{d W_{\text{fric.}}}{dt} = - \dot{x}(t) \int^{\infty}_{-\infty} d\tau \: \gamma(x(t) , t - \tau) \dot{x}(\tau),
\end{align}
where $P_{\text{diss.}} \geq 0$ indicates that the electronic friction has an overall damping effect. In the near-adiabatic limit, the total dissipated energy is determined by the long-time average power at a particular coordinate,
\begin{align}
    \bar{E}_{\text{diss.}}(x) = \: & \int^{\infty}_{-\infty} dt \: P_{\text{diss.}} = \: - \frac{1}{2\pi}\int^{\infty}_{-\infty} d\omega \: \underbrace{|\tilde{\dot{x}}(\omega)|^{2} \tilde{\gamma}(x,\omega)}_{\bar{P}_{\text{diss.}}(x,\omega)}, \label{eq: dissipated power}
\end{align}
where the power density $\bar{P}_{\text{diss.}}(x,\omega)$ has been identified. Exploiting the causality of the electronic frictional kernel, $\gamma(t < 0) = 0$, it can be shown that the power density depends only on the real part,
\begin{align}
    \bar{P}_{\text{diss.}}(x,\omega) = \: & |\tilde{\dot{x}}(\omega)|^{2} \text{Re}\left\{\tilde{\gamma}(x,\omega)\right\}, \label{eq: power density}
\end{align}
such that the electronic friction force has dissipative effect at coordinate $x$ and frequency $\omega$ if $\text{Re}\left\{\gamma(x,\omega)\right\} > 0$, and vice versa for $\text{Re}\left\{\gamma(x,\omega)\right\} < 0$. Since nonequilibrium vibrational dynamics generally involve multiple frequencies, the overall dissipative or driving effect of the electronic friction force cannot be determined from a single frequency component, but instead depends on the full frequency-dependent energy exchange along the vibrational trajectory.

\subsection{Coordinate-Dependent Markovian Embedding} \label{subsec: Coordinate-Dependent Markovian Embedding}

In this section, we introduce the procedure used to propagate the non-Markovian Langevin equation in Eq.\eqref{eq: Generalized Langevin equation}. Although the method is similar to other Markovian embedding approaches \cite{Ceriotti2009_nuclear,Ceriotti2010_colored,Ceriotti2010_efficient,Ceriotti2012_efficient,Trenins_nonmarkovian_2025,Lawrence2019_on,Lu2026_memory} to sampling the colored noise, we note that it differs in several key aspects, namely the violation of the classical FDT and the fact that the coordinate- and time-dependence of the non-Markovian electronic forces does not factorize. 


First, the electronic friction and the correlation function of the stochastic force are represented by a sum over exponential functions:
\begin{align}
\gamma(x(t),t) = \: & \text{Re}\left\{\sum^{N_{\gamma}}_{j=1} a_{j}(x(t)) e^{-b_{j}(x(t))t}\right\} \label{eq: friction decomposition} \\
D(x(t),t) = \: & \text{Re}\left\{ \sum^{N_{D}}_{j=1} c_{j}(x(t)) e^{-d_{j}(x(t))t} \right\}. \label{eq: correlation function decomposition}
\end{align}
For the sake of brevity, from here on we will omit the explicit time dependence of the vibrational coordinate. In general, the weights and frequencies are complex numbers, 
\begin{align}
    a_{j} = \: & a_{j,r} +  i a_{j,i} \:\:\:\: , \:\:\:\: b_{j} = \: b_{j,r} +  i b_{j,i}, \nonumber \\
    c_{j} = \: & c_{j,r} +  i c_{j,i} \:\:\:\: , \:\:\:\: d_{j} = \: d_{j,r} +  i d_{j,i},
\end{align}
although the real components of the frequencies are naturally subject to a positivity constraint
\begin{align}
    b_{j,r}(x),d_{j,r}(x) > 0.
\end{align}
As we later show, the weights of $D(x,t)$ must also satisfy $c_{j}(x) \in \mathbb{R}^{+}$, so that each one represents a physical Ornstein-Uhlenbeck process. 

It is also important to note that, although many Markovian embedding procedures are explicitly applied in equilibrium, where $\gamma(x,t)$ and $D(x,t)$ are related by the classical fluctuation-dissipation theorem in Eq.\eqref{eq: classical fdt} and only one decomposition is required for both $\gamma(x,t)$ and $D(x,t)$ \cite{Ceriotti2009_nuclear,Ceriotti2010_colored,Ceriotti2010_efficient,Ceriotti2012_efficient,Trenins_nonmarkovian_2025,Lawrence2019_on}, our approach is general for nonequilibrium scenarios. In these regimes, the fluctuation-dissipation theorem does not relate $\gamma(x,t)$ and $D(x,t)$ and they must be decomposed independently, which may require different numbers of terms, $N_{D} \neq N_{\gamma}$. We discuss suitable decomposition procedures further in Sec.~\ref{subsec: Coordinate-Dependent Pole Decomposition}.

Next, we define the deterministic auxiliary variables,
\begin{align}
u_{j}(t) = \: \int^{t}_{0} d\tau \: a_{j}(x) e^{-b_{j}(x)(t-\tau)} p(\tau),
\end{align}
and map to a real representation, 
\begin{align}
    u_{j} \longrightarrow \mathbf{u}_{j} = [u_{j,r},u_{j,i}].
\end{align}
The corresponding equation of motion is obtained by differentiating with respect to time,
\begin{align}
    \frac{d \mathbf{u}_{j}}{d t} = \: & \frac{\partial \mathbf{u}_{j}}{\partial t} + \frac{\partial \mathbf{u}_{j}}{\partial x} \dot{x}(t),
\end{align}
where the second term arises from the explicit coordinate dependence of $a_{j}(x)$ and $b_{j}(x)$. Since the generalized Langevin equation has already been derived under the quasi-stationary approximation, $x(\tau) \approx x(t)$, the decomposition coefficients $a_{j}(x)$ and $b_{j}(x)$ are interpreted as slowly varying parameters that remain effectively constant over the electronic relaxation timescale. Consequently, the auxiliary variables inherit this approximation, and their evolution is dominated by the explicit time dependence of the exponential kernel. As a result, the contribution arising from the coordinate dependence, $\frac{\partial \mathbf{u}_{j}}{\partial x} \dot{x}(t)$, represents a higher-order correction in the electronic-vibrational timescale separation and is neglected throughout this work.

Consequently, in the quasi-stationary limit, the time evolution of the deterministic auxiliary variables is given by
\begin{align}
\frac{d \mathbf{u}_{j}}{d t} \approx \: & \frac{\partial \mathbf{u}_{j}}{\partial t} \\
= \: & \mathbf{a}_{j}(x(t)) p(t) - \mathbf{B}_{\gamma,j}(x(t)) \mathbf{u}_{j}(t),
\end{align}
where the generator of the dynamics given by 
\begin{align}
 \mathbf{B}_{\gamma,j}(x(t)) = \: & \left[\begin{array}{c c}
                        b_{j,r} & -b_{j,i} \\
                        b_{j,i} & b_{j,r}
                        \end{array}\right].
\end{align}
Since the electronic forces are calculated in the limit of a timescale separation between electronic and vibrational degrees of freedom, it is natural to initialize the auxiliary variables in their stationary state, $\dot{\mathbf{u}}_{j}(0) = 0$, which leads to the initial condition
\begin{align}
    \mathbf{u}_{j}(0) = [\mathbf{B}_{\gamma,j}(x_{0})]^{-1}\mathbf{a}_{j}(x_{0}) p_{0},
\end{align}
where $\{x_{0},p_{0}\} = \{x(0),p(0)\}$ is the vibrational initial condition. Consequently, the electronic friction force then takes the form
\begin{align}
    \int^{t}_{0} d\tau \: \gamma(x(t),t - \tau) p(\tau) = \sum_{j = 1}^{N_{\gamma}} u_{j,r}(t).
\end{align}

Next, a similar approach is employed to represent the stochastic force, which is reconstructed as a sum over independently sampled Ornstein-Uhlenbeck (OU) processes,
\begin{align}
f(t) = \sum_{j = 1}^{N_{D}} w_{j,r}(t), \label{eq: reconstructed stochastic force}
\end{align}
where $w_{j,r}(t)$ is the real component of a complex stochastic variable,
\begin{align}
    w_{j} \longrightarrow \boldsymbol{w}_{j} = [w_{j,r},w_{j,i}].
\end{align}
The $\boldsymbol{w}_{j}$ are referred to as auxiliary stochastic variables and evolve according to the Ornstein-Uhlenbeck equation of motion,
\begin{align}
d\boldsymbol{w}_{j}(t) = \: & - \mathbf{B}_{D,j}(x(t))\boldsymbol{w}_{j}(t)dt
+ \boldsymbol{\sigma}_{j}(x(t))\circ d\mathbf{W}_{t},
\end{align}
where the deterministic generator of the $j$th OU process is
\begin{align}
\mathbf{B}_{D,j}(x(t)) = \: &
\left[
\begin{array}{c c}
d_{j,r}(x(t)) & -d_{j,i}(x(t))\\
d_{j,i}(x(t)) & d_{j,r}(x(t))
\end{array}
\right].
\end{align}
Here, $d\mathbf{W}_{t}=[dW_{r,t},dW_{i,t}]$ denotes a vector of independent Wiener processes, while $\boldsymbol{\sigma}_{j}(x(t))$ is the corresponding noise amplitude of the $j$th OU process. Consistent with the deterministic auxiliary variables and the quasi-stationary approximation, the coordinate dependence of the OU processes is treated locally along the vibrational trajectory. Specifically, the vibrational coordinate is assumed to remain approximately constant over the electronic correlation time $\tau_{\text{el}}$, such that the stochastic variables relax under an instantaneous electronic environment defined by $x(t)$. Consequently, each OU process represents a locally stationary stochastic process, with its covariance determined by the stationary Lyapunov equation associated with $\mathbf{B}_{D,j}(x)$ and $\boldsymbol{\sigma}_{j}(x)$. The parameters of the OU processes are subsequently updated according to the instantaneous vibrational coordinate during propagation.

If the weights of the correlation-function decomposition satisfy $c_{j}(x)\in\mathbb{R}^{+}$, then the noise amplitudes can be chosen as 
\begin{align}
    \boldsymbol{\sigma}_{j}(x) = \sqrt{2 \: d_{j,r}(x) \: c_{j}(x)}\mathbb{I}_{2},
\end{align}
Consequently, if the OU processes are initialized in their local stationary state, such that the initial conditions are sampled according to
\begin{align}
    \boldsymbol{w}_{j}(0) \sim \mathcal{N}\left(\boldsymbol{0},c_{j}(x_{0})\mathbb{I}_{2}\right),
\end{align}
then the covariance of the real components is given by
\begin{align}
\langle w_{j,r}(t)w_{j',r}(\tau)\rangle = \: & \delta_{jj'}\text{Re}\left\{c_{j}(x(t))e^{-d_{j}(x(t))(t-\tau)}\right\},
\end{align}
As a result, the reconstructed stochastic force in Eq.\eqref{eq: reconstructed stochastic force} reproduces the target electronic force correlation function, which is shown in detail in the Suppplementary Material. 

The original non-Markovian Langevin equation in Eq.\eqref{eq: Generalized Langevin equation} has now been mapped to a series of coupled Markovian Langevin equations, where the influence of the friction and stochastic forces are incorporated via deterministic and stochastic auxiliary variables, respectively. The final coupled equations of motion are
\begin{align}
\dot{x} = \: & p / m \qquad \nonumber \\
\dot{p} = \: & - \partial_{x} U_{\text{vib}} + F^{\text{ad}}_{\text{el}}(x) - \frac{1}{m}\sum_{j}^{N_{\gamma}} u_{j,r}(t) + \sum_{k}^{N_{D}} w_{j,r}(t) \nonumber  \\
\dot{\mathbf{u}}_{j}(t) = \: & \mathbf{a}_{j} p(t) - \mathbf{B}_{\gamma,j}(x(t)) \mathbf{u}_{j}(t) \nonumber \\
d\boldsymbol{w}_{j}(t) = \: & - \mathbf{B}_{D,j}(x(t)) \boldsymbol{w}_{j}(t) dt + \boldsymbol{\sigma}_{j} (x(t)) \circ d\mathbf{W}_{t}, \label{eq: coupled Markovian LEs to solve}
\end{align}
with initial conditions 
\begin{align}
    \{x(0),p(0)\} = \: & \{x_{0},p_{0}\} \label{eq: phys initial conditions} \\
    \mathbf{u}_{j}(0) = \: & [\mathbf{B}_{\gamma,j}(x_{0})]^{-1}\mathbf{a}_{j}(x_{0}) p_{0} \label{eq: det aux initial conditions} \\
    \boldsymbol{w}_{j}(0) \sim \: & \mathcal{N}\left(\boldsymbol{0},c_{j}(x_{0})\mathbb{I}_{2}\right). \label{eq: stoch aux initial conditions}
\end{align}

\section{NM-EFLD Algorithm} \label{sec: NM-EFLD Algorithm}

In this section, we provide a detailed outline of the procedure used to solve the Markovian-embedded equations of motion in Eq.\eqref{eq: coupled Markovian LEs to solve}. This involves an outline of the general algorithm in Sec.~\ref{subsec: General Algorithm}, in which the sum-over-exponentials decomposition procedure and the updated ABOBA propagation are discussed, as well as a discussion of observables and numerical details in Sec.~\ref{subsec: Observables and Numerical Details}. 

\subsection{General Algorithm} \label{subsec: General Algorithm}

The general workflow can be split into three parts:
\begin{enumerate}
    
    \item First, the electronic forces, comprising $F^{\text{ad}}_{\text{el}}(x)$, $\gamma(x,t)$, and $D(x,t)$, are obtained by an appropriate quantum dynamical method. In this work, the HEOM approach was used to compute the electronic forces \cite{Rudge2023,Maeck2024}, with exact expressions relegated to Sec.~V of the Supplementary Material. 

    \item The electronic forces are preprocessed to prepare them for simulation in the NM-EFLD approach. The adiabatic contribution to the mean electronic force is interpolated, while the electronic friction and correlation function of the stochastic force are decomposed into sums over exponential functions, as per Eq.\eqref{eq: friction decomposition} and Eq.\eqref{eq: correlation function decomposition}. To obtain coordinate-dependent weights and frequencies, this decomposition is performed for each point of the relevant $x$-coordinate range and the resulting weights and frequencies are interpolated. More details on this procedure are given in Sec.~II of the Supplementary Material.

    \item In the final part of the procedure, the decomposed electronic forces are used in Eq.\eqref{eq: coupled Markovian LEs to solve} to propagate the coupled Markovian equations of motion. To do this, we employ an updated ABOBA algorithm, which is itself a second-order symmetric decomposition of the stochastic propagator. Specifically, we symmetrically split the momentum update further to accommodate the time evolution of the deterministic and stochastic auxiliary variables. Further details are given in Sec.~III of the Supplementary Material, while details on the initial conditions, observables, and ensemble averaging are given in Sec.~\ref{subsec: Observables and Numerical Details}.
    
\end{enumerate}

Note that the preprocessing in Step $2$ does not have to occur before the propagation starts. In principle, one could design an on-the-fly algorithm, where the exponential decomposition is performed during the propagation at the current vibrational coordinate. For one-dimensional systems, it was found that preprocessing the non-Markovian forces was highly efficient. However, in higher-dimensional systems with multiple vibrational modes, the memory required to store the decomposed forces grows exponentially, and it may be more efficient to decompose only the current $x$-coordinate rather than the entire grid. 

\subsection{Observables and Numerical Details} \label{subsec: Observables and Numerical Details}

We first note that the unperturbed vibrational potential used in this work, $U_{\text{vib}}$, is a harmonic oscillator, as shown in Eq.\eqref{eq: molecular Hamiltonian donor-acceptor}. Consequently, we perform all simulations in dimensionless coordinates,
\begin{align}
    \{x,p\} \longrightarrow \{x\sqrt{m\Omega},p/\sqrt{m\Omega}\}.
\end{align} 
The vibrational initial conditions are sampled from the Wigner function of the ground state of the unperturbed harmonic oscillator, $\{x_{0},p_{0}\} \sim \rho_{\text{W}}(x,p)$, with 
\begin{align}
    \rho_{\text{W}}(x,p) = \: & \frac{1}{\pi} e^{-(x^{2} + p^{2})}.
\end{align}

Once the physical initial conditions are sampled, they are used to construct the initial conditions of the deterministic and auxiliary variables in Eq.\eqref{eq: det aux initial conditions} and Eq.\eqref{eq: stoch aux initial conditions}. During propagation, the ensemble of stochastic trajectories can be sampled to obtain expectation values of observables, with vibrational observables specifically forming the main focus of this work. For example, the average kinetic and potential energies are given by 
\begin{align}
    \langle \text{KE} \rangle(t) = \: & \frac{1}{N_{\text{traj.}}}\sum_{k = 1}^{N_{\text{traj.}}} \frac{\Omega}{2} p^{2}_{t,k}, \\
    \langle \text{PE} \rangle(t) = \: & \frac{1}{N_{\text{traj.}}}\sum_{k = 1}^{N_{\text{traj.}}} \frac{\Omega}{2} x^{2}_{t,k}, \\
    \langle E_{\text{vib}} \rangle(t) = \: & \langle \text{KE} \rangle(t) + \langle \text{PE} \rangle(t).
\end{align}

In the same manner, the simulated correlation function of the stochastic force can be compared with the theoretical prediction. In the quasi-stationary approximation, the electronic correlation function is evaluated at a fixed vibrational coordinate. Therefore, for an ensemble of trajectories initialized with different nuclear coordinates, the corresponding theoretical correlation function is obtained by averaging over the initial conditions,
\begin{align}
    \langle D(x_{0},t) \rangle = \: & \frac{1}{N_{\text{traj.}}}\sum_{k=1}^{N_{\text{traj.}}} D(x_{0,k},t), \label{eq: IC corr func average}
\end{align}
where $x_{0,k}$ denotes the initial coordinate of the $k$th trajectory. This comparison assumes that the stochastic force is initialized in its stationary distribution, as ensured by the auxiliary variable initial conditions in Eq.\eqref{eq: stoch aux initial conditions}. The corresponding correlation function obtained from the stochastic trajectories is calculated relative to the initial time of each trajectory,
\begin{align}
    \langle f(t)f(0)\rangle = \: & \frac{1}{N_{\text{traj.}}} \sum_{k=1}^{N_{\text{traj.}}}    \left(f_{t,k}-\bar{f}_{k}\right)\left(f_{0,k}-\bar{f}_{k}\right), \label{eq: sampled IC corr func}
\end{align}
where $\bar{f}_{k}$ is the mean stochastic force along the $k$th trajectory, subtracted to remove finite-sampling bias in the numerical estimate.

\begin{table}
    \begin{tabular}{|| c | c | c | c ||}
    \hline
        $\Omega$ & $k_{B}T$ & $\Gamma$ & $\lambda$ \\ 
    \hline 
        $30\text{meV}$ & $25.85\text{meV}$ & $100\text{meV}$ & $10\text{meV}$\\
    \hline
    \end{tabular}
    \caption{Parameters for the vibronic model investigated in this work. The energy separation between the two electronic states, $2\Delta$, is varied between investigations.}
    \label{tbl: parameters}
\end{table}

To calculate expectation values of vibrational observables in the nonequilibrium stationary state as a function of bias voltage, we utilize a separate procedure. Namely, the trajectory ensemble is propagated to a time $t_{\text{ss}}$, at which point the average vibrational energy stops evolving,
\begin{align}
    \frac{d}{dt}\langle E_{\text{vib}} \rangle \Big |_{t_{\text{ss}}} = \: & 0.
\end{align}
From here, we propagate each trajectory further and sample $N_{\text{sample}}$ points from each, with each point separated by an equilibration time $t_{\text{ind.}}$ to ensure sampling independence, for a total of $N_{\text{tot.}} = N_{\text{sample}} \times N_{\text{traj.}}$ points:
\begin{align}
    \langle E_{\text{vib}} \rangle^{\text{ss}} = \: & \frac{1}{N_{\text{tot.}}}\sum_{k = 1}^{N_{\text{tot.}}} \frac{\Omega}{2} \left(x_{k}^{2} + p_{k}^{2}\right).
\end{align}
From this, the average vibrational excitation number in the steady state can be estimated by 
\begin{align}
    \langle N_{\text{vib}}\rangle^{\text{ss}} = \: & \langle E_{\text{vib}}\rangle^{\text{ss}} / \Omega.
\end{align}
All steady-state expectation values of observables were calculated from $N_{\text{traj.}} = 500$ trajectories with \newline $N_{\text{sample}} = 500$ samples from each and an independence time of $t_{\text{ind.}} = 3 \Omega^{-1}$.

Note that, although vibrational observables form the main focus of this work, the procedure outlined here can also be applied to obtain expectation values of electronic observables, as shown in detail in Refs.~\cite{Rudge2024,Maeck_Vibrational2025,Preston2022}. 

\begin{figure}
    \centering
    \includegraphics[width=\linewidth]{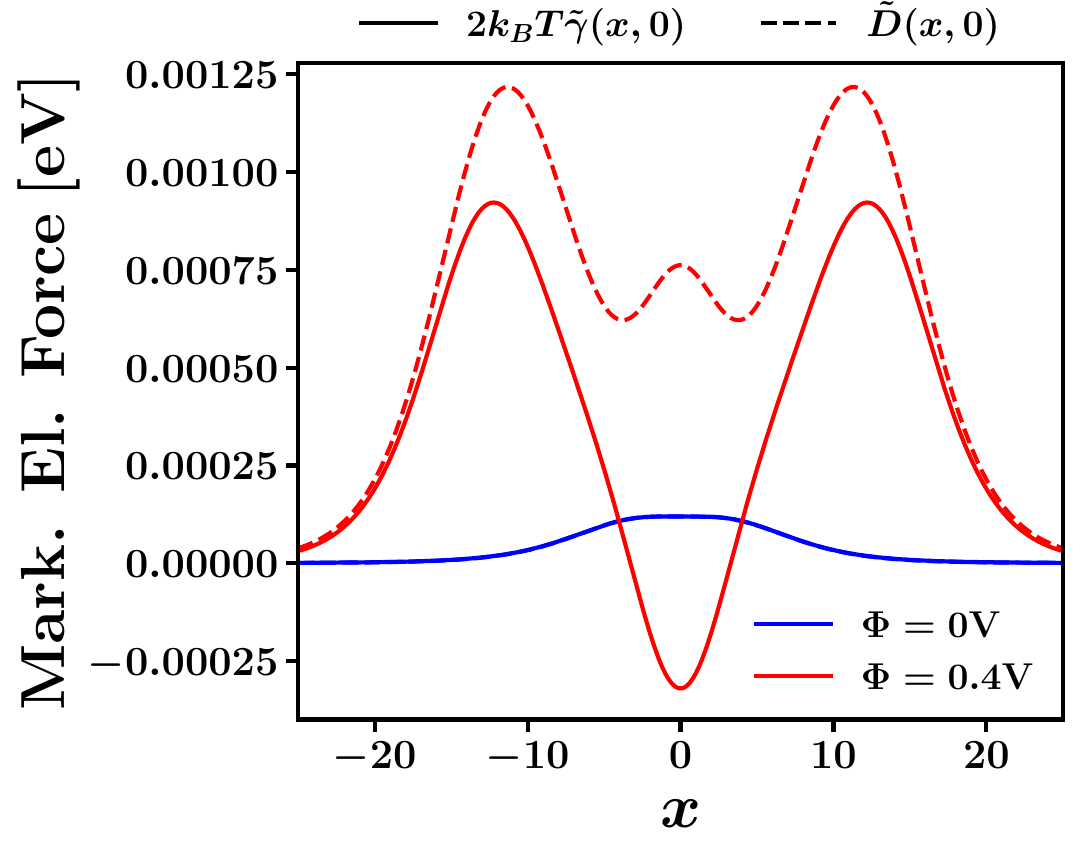}
    \caption{Markovian electronic friction (solid lines) and correlation function of the stochastic force (dashed lines) plotted as a function of vibrational coordinate, $x$, and for two voltages: $\Phi = 0\text{V}$ (blue) and $\Phi = 0.4\text{V}$ (red). The electronic energies are chosen as $\Delta = 100\text{meV}$ and all other parameters are given in Tbl.~\ref{tbl: parameters}.}
    \label{fig: markovian electronic forces pos delta}
\end{figure}

\section{Results} \label{sec: Results}

In this section, we present the non-Markovian electronic forces, their exponential decompositions, and simulations of the resulting generalized Langevin equation for an illustrative vibronic model system. These results are compared against numerically exact fully quantum simulations obtained using the HEOM approach \cite{Tanimura1989,Tanimura2006,Tanimura2020,Haertle2015,Jin2007,Jin2008,Schinabeck2018,Zheng2009,Yan2014,Wenderoth2016,Ye2016}, an outline of which is given in the Supplementary Material.

Specifically, we consider nonequilibrium charge transport through the vibrationally coupled donor-acceptor model introduced in Eq.~\eqref{eq: molecular Hamiltonian donor-acceptor}. While this model has been extensively studied using a range of theoretical approaches \cite{simine_vibrational_2012,segal_heat_2005,segal_heat_2006,segal_nonequilibrium_2011,Foti2018_origin,foti_interface_2017,Lue2011}, recent work has highlighted significant non-Markovian contributions to the electronic forces \cite{Preston2026_negative}. Such memory effects may strongly influence the resulting vibrational dynamics, particularly through their connection to nonequilibrium phenomena such as negative electronic friction and current-induced heating. Although the present calculations focus on a single model system, the non-Markovian Langevin framework and the proposed numerical approach are applicable to a broad class of molecular systems coupled to metallic environments. 

\subsection{Coordinate-Dependent Pole Decomposition} \label{subsec: Coordinate-Dependent Pole Decomposition}

\begin{figure}
    \centering
    \phantomsubfloat{\label{fig: electronic force spectrum x = 0 (a)}}
    \phantomsubfloat{\label{fig: electronic force spectrum x = 0 (b)}}
    \includegraphics[width=\linewidth]{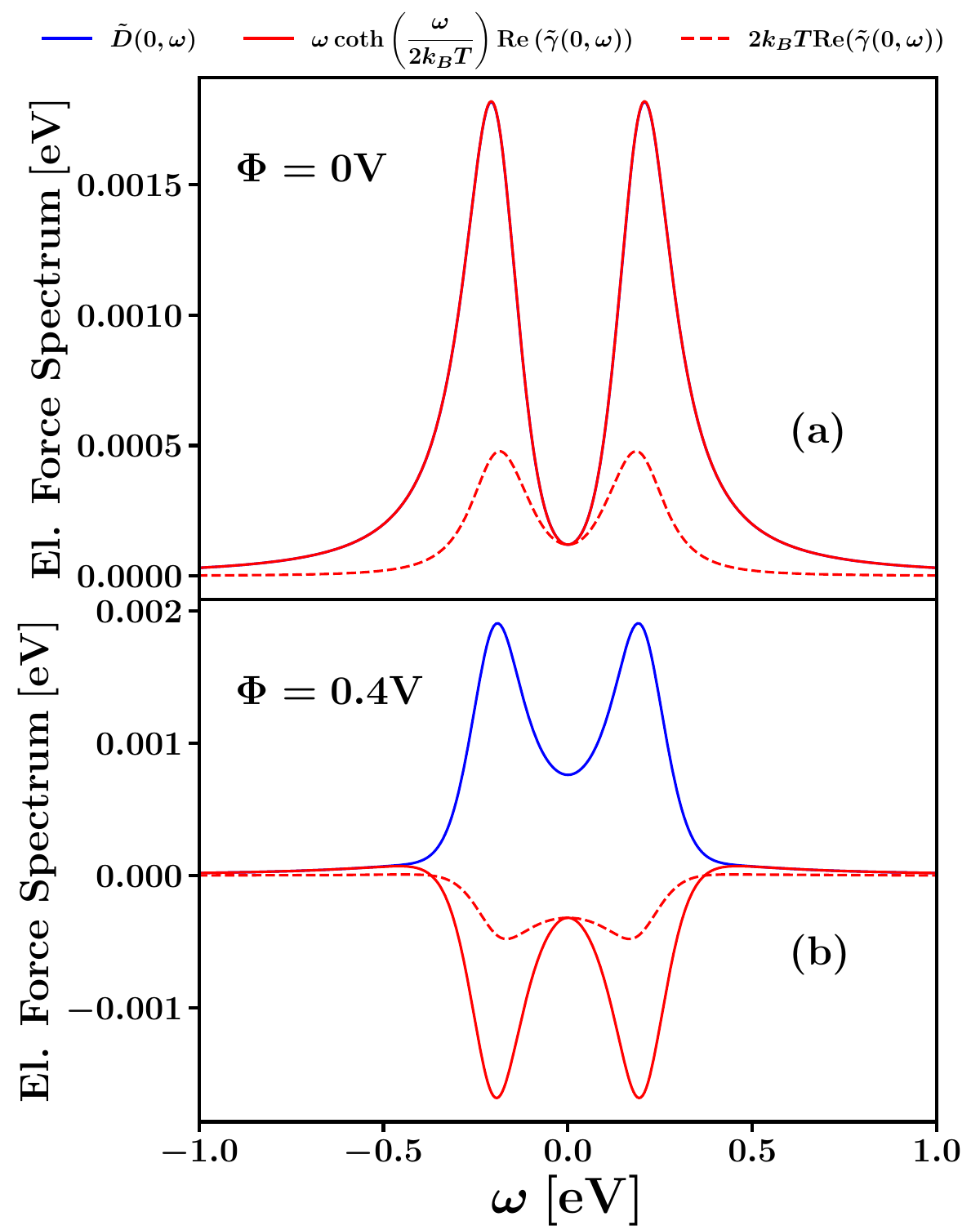}
    \caption{Frequency spectrum of the non-Markovian electronic forces at vibrational coordinate $x = 0$ and for voltages (a) $\Phi = 0\text{V}$ and (b) $\Phi = 0.4\text{V}$. The electronic energies are chosen as $\Delta = 100\text{ meV}$. In order to compare to the correlation function of the stochastic force (blue), the electronic friction (red) has been scaled by the quantum (solid lines) and classical (dashed lines) fluctuation-dissipation theorem factors.}
    \label{fig: electronic force spectrum x = 0}
\end{figure}

In this subsection, we demonstrate the pole decomposition methods described in Sec.~\ref{subsec: General Algorithm} using non-Markovian electronic forces of the vibrationally coupled donor-acceptor model, which were obtained via the HEOM approach \cite{Rudge2023,Maeck2024}. We first apply the method to electronic forces at equilibrium, $\Phi = 0$, where the quantum fluctuation-dissipation theorem is satisfied, and then to a nonequilibrium scenario, $\Phi = 0.4\text{V}$, which highlights the importance of the independent pole decompositions for $\gamma(x,t)$ and $D(x,t)$. For these and all later results, we use the parameters given in Tbl.~\ref{tbl: parameters}, with this subsection exclusively focusing on the $\Delta = 100\text{ meV}$ case of the energetic orientations. 

\begin{figure}
    \centering
    \includegraphics[width=\linewidth]{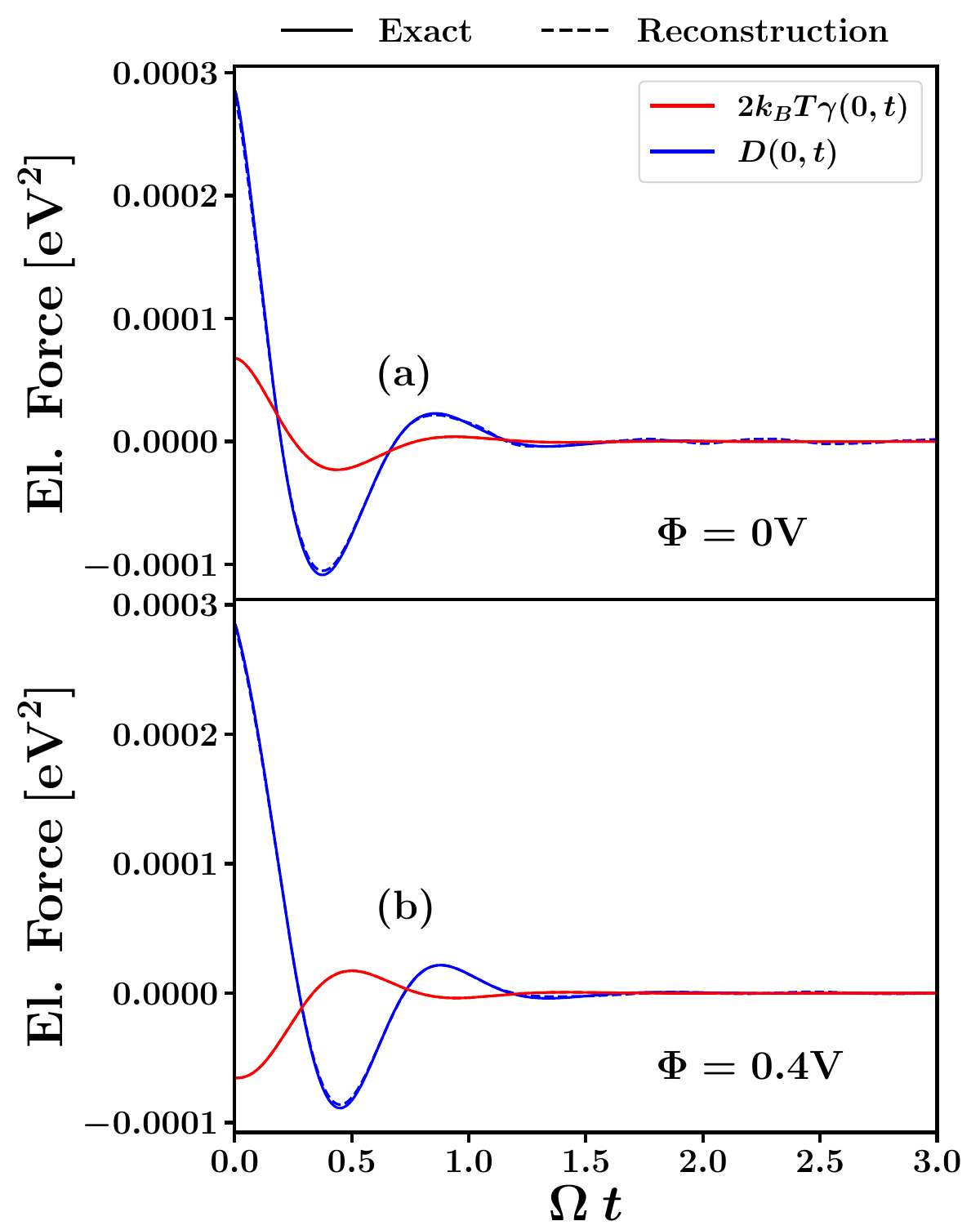}
    \caption{Exact (solid) and reconstructed (dashed) time-dependent non-Markovian electronic forces at vibrational coordinate $x = 0$ and for voltages (a) $\Phi = 0\text{V}$ and (b) $\Phi = 0.4\text{V}$. Parameters are the same as in Fig.~\figref{fig: electronic force spectrum x = 0}. $N_{\gamma} = 6$ terms were used to represent $\gamma(x,t)$, while $N_{D} = 20$ terms were required for an accurate representation of $D(x,t)$.}
    \label{fig: electronic force time x = 0}
\end{figure}

First, we examine the Markovian limit of the electronic forces in Fig.~\figref{fig: markovian electronic forces pos delta}. In equilibrium at zero bias voltage, the electronic friction and stochastic force satisfy the classical fluctuation-dissipation theorem in Eq.\eqref{eq: classical fdt}. In contrast, under nonequilibrium transport conditions, this relation is no longer enforced. Specifically, for finite bias voltage, the stochastic force strength exceeds the electronic friction over a significant vibrational coordinate region, $\tilde{D}(x,0)> 2k_{B}T\tilde{\gamma}(x,0)$, corresponding to current-induced stochastic vibrational heating. 

Furthermore, the Markovian electronic friction becomes negative around $x=0$. As discussed in Ref.~\cite{Preston2026_negative}, this negative friction is a signature of deterministic vibrational heating induced by the vibrationally coupled inelastic electronic transitions with $\Delta > 0$, which are biased due to the nonequilibrium electric current. Physically, electrons transporting across the junction must give up energy of $2\Delta$ to the vibrational degree freedom. For $\Delta = 0$, this contribution vanishes, whereas $\Delta < 0$ yields positive friction associated with vibrational damping by the electronic current. An example of this is shown later in Fig.~\figref{fig: markovian electronic forces neg delta} at even larger bias voltage, $\Phi = 0.7\text{V}$.

Beyond the Markovian limit, it was also shown in Ref.~\cite{Preston2026_negative} that these regimes of negative friction are associated with strong non-Markovian features in the electronic forces. To illustrate this, Fig.~\figref{fig: electronic force spectrum x = 0} shows the frequency-dependent electronic friction and stochastic force correlation function, $\tilde{\gamma}(x,\omega)$ and $\tilde{D}(x,\omega)$, at the representative coordinate $x=0$. At equilibrium, Fig.~\figref{fig: electronic force spectrum x = 0 (a)}, the finite-frequency quantities satisfy the quantum fluctuation-dissipation theorem, while the classical fluctuation-dissipation relation is recovered only in the Markovian limit, $\omega=0$. 

At finite bias voltage, however, even the quantum fluctuation-dissipation relation is violated. For $\Phi=0.4\text{V}$, the electronic friction remains negative across the full frequency range, while the stochastic force correlation function remains positive. Importantly, the finite-frequency peaks in both $\tilde{\gamma}(0,\omega)$ and $\tilde{D}(0,\omega)$ are comparable to or larger than their Markovian values at $\omega=0$, demonstrating that the zero-frequency approximation may neglect significant dynamical structure in the electronic response. Indeed, analysis over the full coordinate range reveals regions where $\tilde{\gamma}(x,0)$ has the opposite sign to $\tilde{\gamma}(x,\omega\neq0)$, indicating that the Markovian friction coefficient can qualitatively misrepresent the underlying electronic back-action. This is discussed in further detail in Sec.~\ref{subsec: Transient and Steady-State Observables}. 

Next, in Fig.~\figref{fig: electronic force time x = 0}, we plot the corresponding electronic forces in the time domain. Furthermore, we have included not only the exact electronic forces obtained from HEOM calculations, but also the reconstructions obtained via the pole-decomposition methods outlined in Sec.~\ref{subsec: General Algorithm} and the Supplementary Material. To obtain the accuracy demonstrated in this plot, $N_{\gamma} = 6$ exponential terms were required to represent the electronic friction, and $N_{D} = 20$ terms to represent the correlation function. In general, the electronic friction requires fewer poles in the decomposition, as it does not constrain the weights to be real. Fig.~\figref{fig: electronic force time x = 0} also demonstrates that independent pole decompositions are required for $\gamma(x,t)$ and $D(x,t)$, as the two quantities are not related by the simple classical fluctuation-dissipation theorem, even in equilibrium.

\begin{figure}
    \centering
    \includegraphics[width=\linewidth]{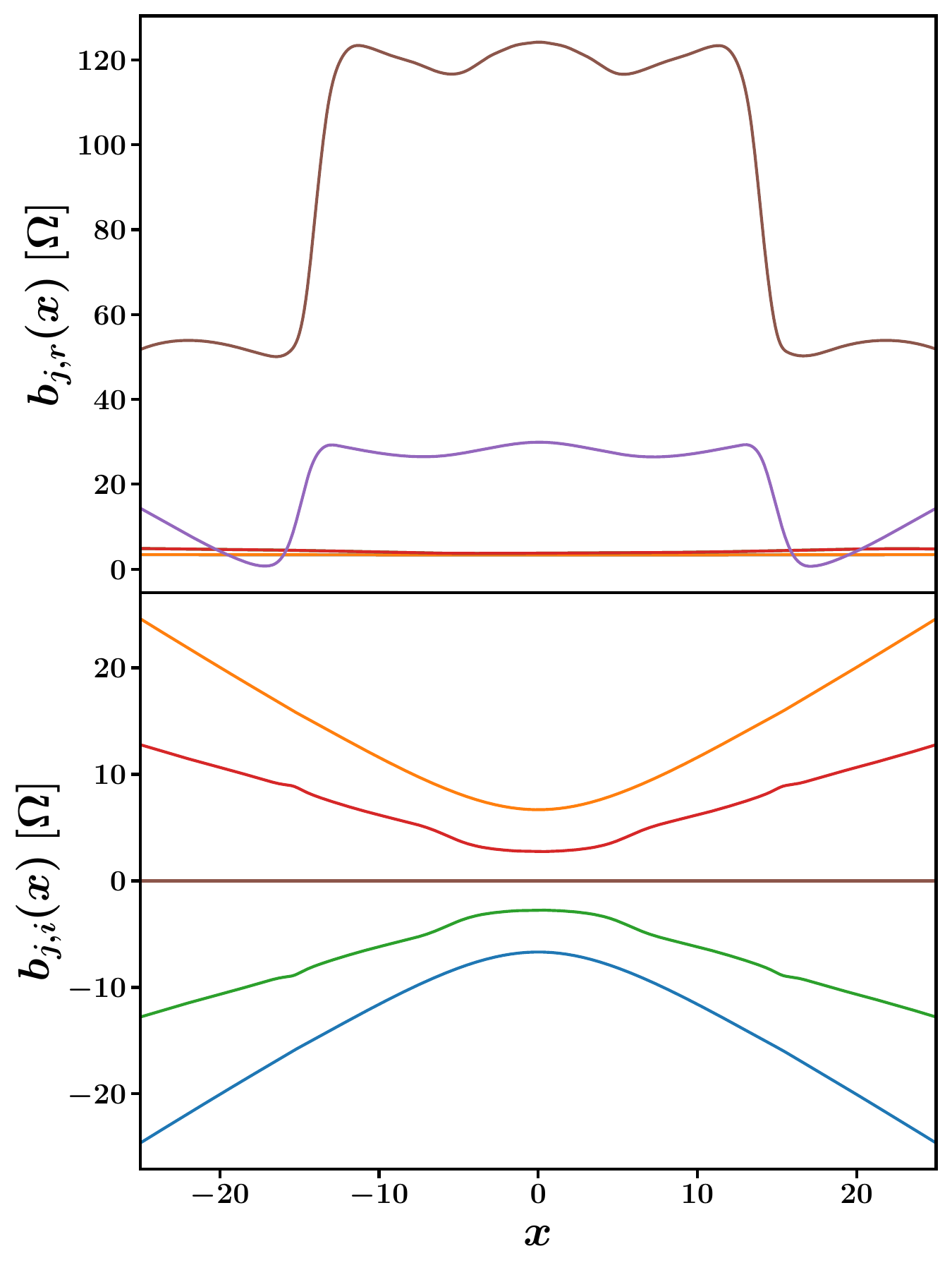}
    \caption{Coordinate-dependent frequencies of the electronic friction, $b_{j}(x)$, obtained by performing the pole-decomposition approach outlined in Sec.~\ref{subsec: General Algorithm} to each vibrational coordinate at $\Phi = 0\text{V}$. The varying colors denote the $N_{\gamma} = 6$ terms used in this pole decomposition. The frequencies are split into real (upper) and imaginary (lower) components.}
    \label{fig: electronic friction frequencies phi = 0}
\end{figure}

Finally in this subsection, the pole decomposition procedure is applied to each vibrational coordinate in Fig.~\figref{fig: markovian electronic forces pos delta}, generating the coordinate-dependent weights and frequencies for both the electronic friction and the correlation function of the stochastic force. As an example, the frequencies of the electronic friction at $\Phi = 0\text{V}$ are shown in Fig.~\figref{fig: electronic friction frequencies phi = 0}. As required by the non-Markovian Langevin theory, the frequencies evolve smoothly with $x$. The corresponding weights, as well as the decomposition of the correlation function, are shown in the Supplementary Material.

\subsection{Transient and Steady-State Observables} \label{subsec: Transient and Steady-State Observables}

\begin{figure}
    \centering
    \includegraphics[width=\linewidth]{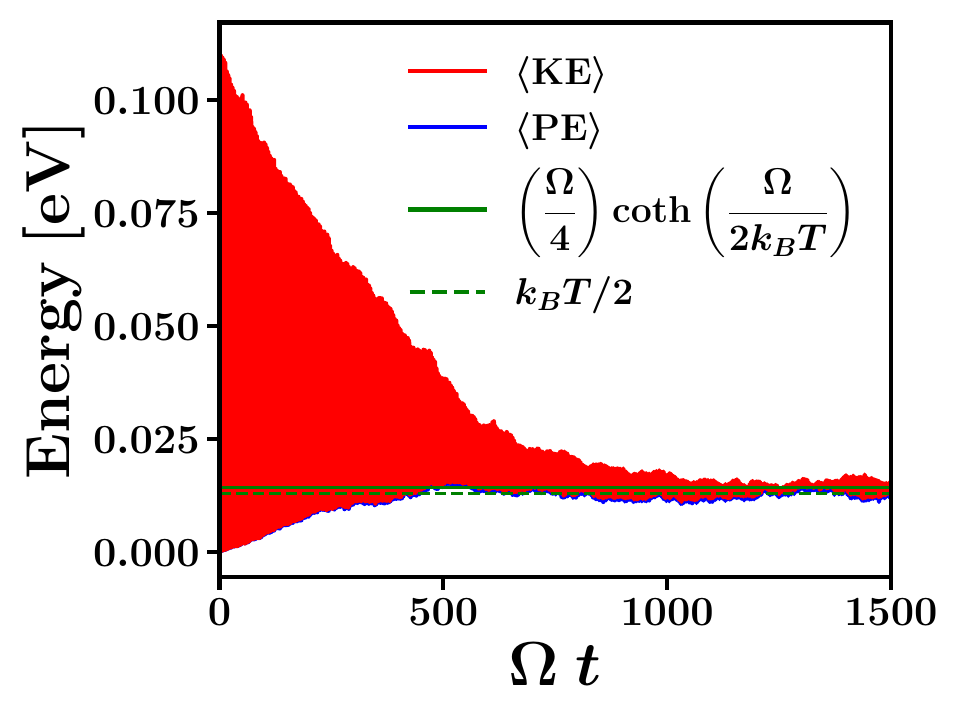}
    \caption{Time-dependent average vibrational kinetic and potential energy at $\Phi = 0\text{ V}$ and for $\Delta = 100\text{ meV}$. Also shown in green are the classical equipartition theorem value (dashed), $k_{B}T / 2$, and the quantum thermal expectation value (solid), $\frac{\Omega}{4}\coth\left(\frac{\Omega}{2k_{B}T}\right)$. A total of $N_{\text{traj.}} = 100$ trajectories were used.}
    \label{fig: equipartition theorem check}
\end{figure}

In this subsection, we employ the pole decompositions introduced above to investigate the time-dependent vibrational dynamics of the vibrationally coupled donor-acceptor model and simulated with the NM-EFLD approach. This approach retains the full frequency dependence of the electronic friction and stochastic force correlation functions, allowing us to directly assess the impact of non-Markovian electronic back-action on the vibrational dynamics. 

\begin{figure}
    \centering
    \includegraphics[width=\linewidth]{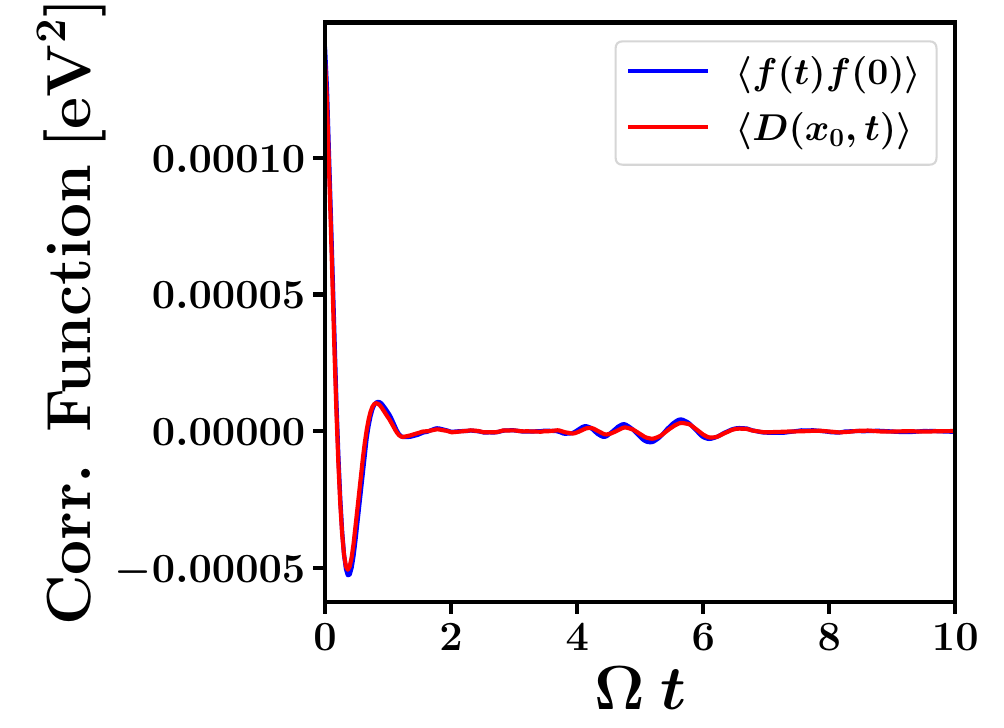}
    \caption{Exact (red) and estimated (blue) average correlation function at $\Phi = 0\text{V}$. Parameters are the same as in Fig.~\figref{fig: equipartition theorem check}.}
    \label{fig: correlation function zero bias}
\end{figure}

In Fig.~\figref{fig: equipartition theorem check}, we first examine the equilibrium vibrational dynamics at zero bias voltage, $\Phi=0\text{V}$. In the absence of nonequilibrium electronic driving, the average kinetic and potential energies, $\langle \text{KE}\rangle(t)$ and $\langle \text{PE}\rangle(t)$, relax to their stationary values of $\frac{\Omega}{4}\coth\left(\frac{\Omega}{2k_{B}T}\right)$, consistent with a harmonic oscillator in thermal equilibrium with a quantum bath at temperature $T$. Note that these energies do not relax to the classical equipartition theorem value of $k_{B}T/2$, which contrasts equilibrium statistics obtained from the M-EFLD approach \cite{Rudge2024}. As a further validation of the stochastic force sampling, Fig.~\figref{fig: correlation function zero bias} compares the correlation function obtained from $N_{\text{traj.}}=100$ trajectories with the exact correlation function averaged over the initial conditions, as outlined in Eq.\eqref{eq: IC corr func average} and Eq.\eqref{eq: sampled IC corr func}. The excellent agreement demonstrates that the auxiliary-variable representation accurately reproduces the statistics of the non-Markovian stochastic force.

\begin{figure}
    \centering
    \includegraphics[width=\linewidth]{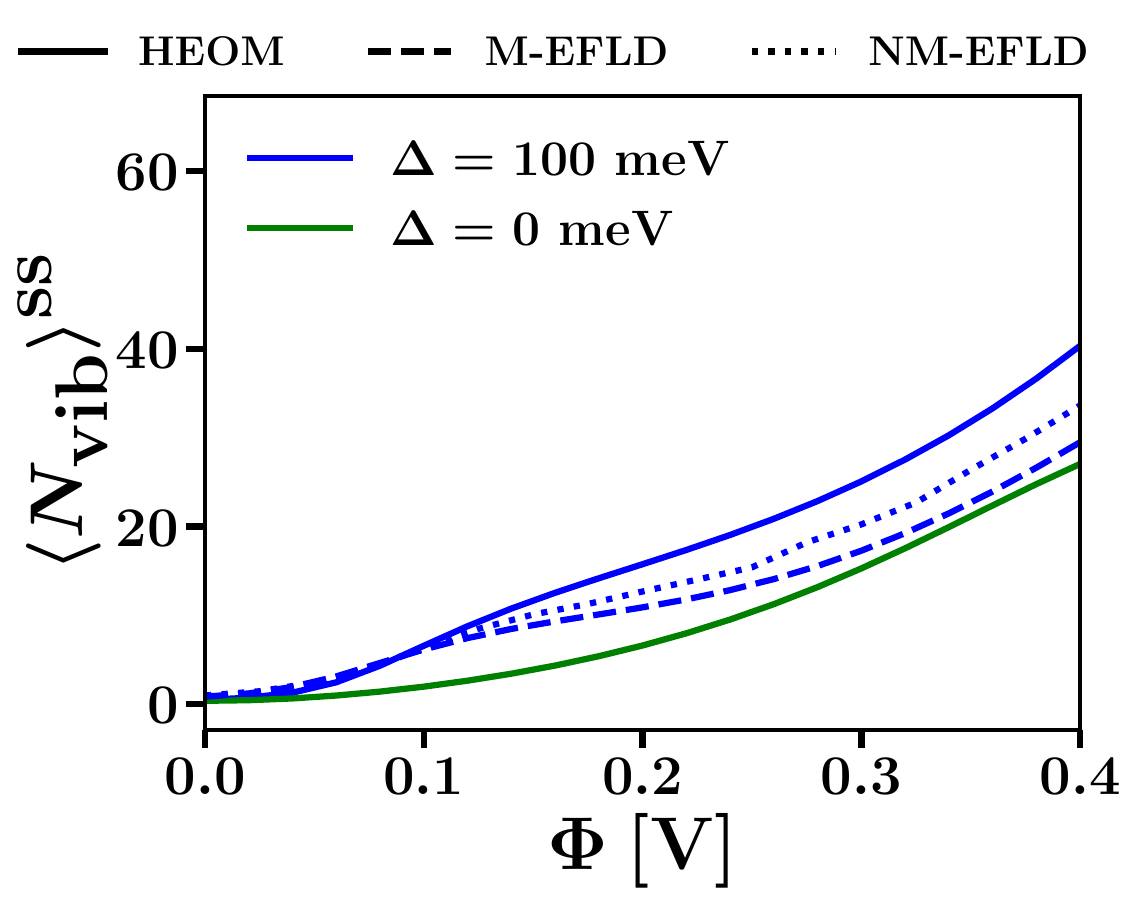}
    \caption{Steady-state vibrational excitation as a function of the bias voltage for two different energetic configurations and simulated via three different methods. The blue (green) lines refer to $\Delta = 100\text{ meV}$ ($\Delta = 0\text{ meV}$) while the solid, dashed, and dotted lines refer to the fully quantum HEOM approach, the Markovian EFLD approach, and the NM-EFLD approach, respectively.}
    \label{fig: pos Delta NESS results}
\end{figure}

\begin{figure}
    \centering
    \phantomsubfloat{\label{fig: neg Delta NESS results (a)}}
    \phantomsubfloat{\label{fig: neg Delta NESS results (b)}}
    \includegraphics[width=\linewidth]{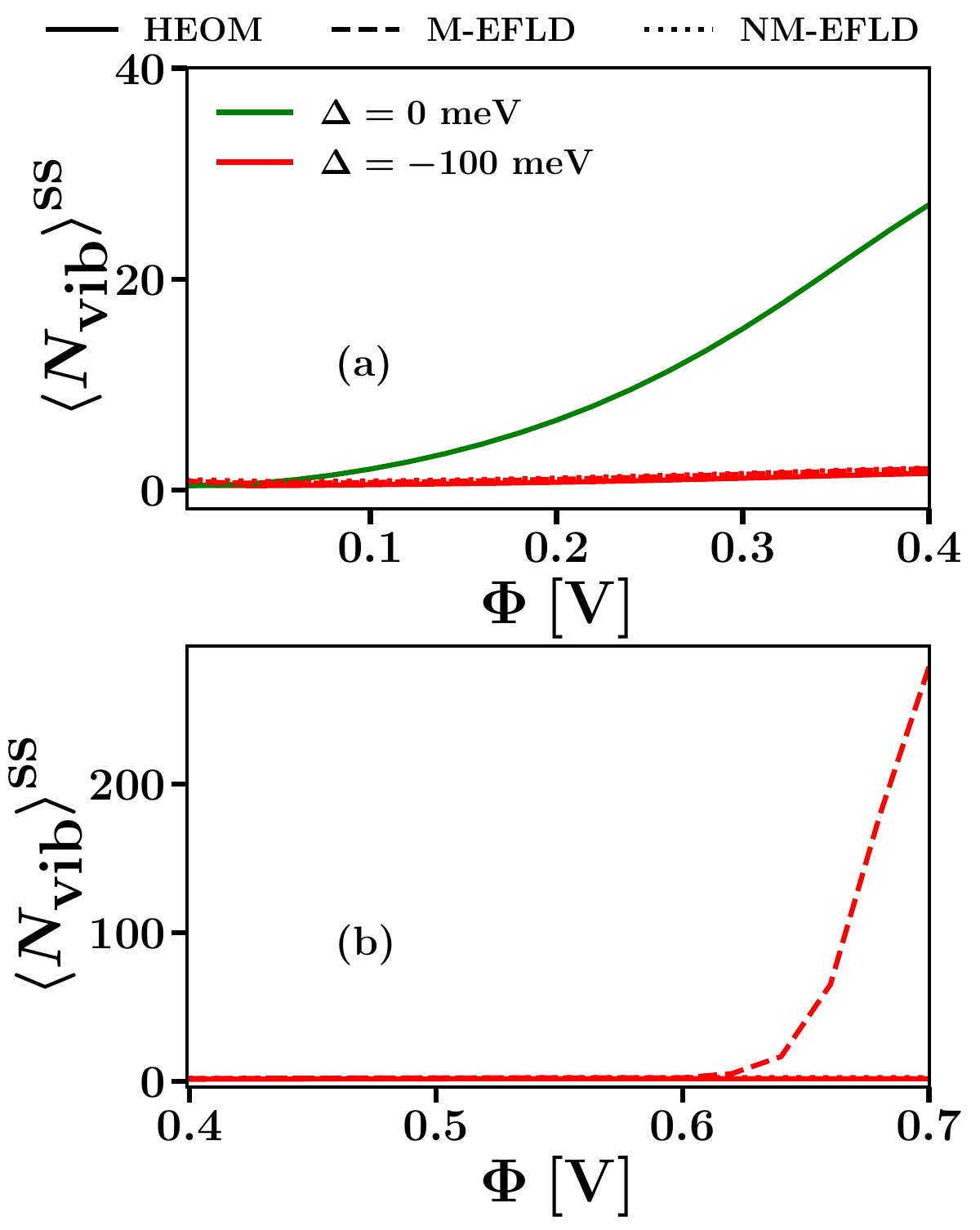}
    \caption{Steady-state vibrational excitation as a function of the bias voltage for $\Delta = -100\text{ meV}$ (red) and $\Delta = 0\text{ meV}$ (green). In (a), the voltage range up to $\Phi = 0.4\text{ V}$ is shown, where the M-EFLD remains stable, while (b) shows the range $\Phi = 0.4-0.7\text{ V}$, where the M-EFLD grows unstable for $\Delta = -100\text{ meV}$. Note that due to numerical challenges, purely quantum HEOM results for $\Delta = 0\text{ meV}$ are only available up to $\Phi = 0.4\text{ V}$.}
    \label{fig: neg Delta NESS results}
\end{figure}

While these results establish the physical consistency of the NM-EFLD approach, they do not yet reveal the consequences of retaining the full non-Markovian structure of the electronic forces. Indeed, at equilibrium, the Markovian EFLD approach also reproduces the correct stationary vibrational temperature through the fluctuation-dissipation theorem, meaning that thermodynamic observables alone cannot distinguish between Markovian and non-Markovian descriptions. The impact of electronic memory is therefore expected to emerge primarily in regimes where the detailed time dependence of the electronic back-action is important, including transient processes such as scattering and desorption dynamics, as well as the nonequilibrium stationary state driven by finite bias voltage.

This question was the central motivation of Ref.~\cite{Preston2026_negative}, where the presence of additional electronic timescales in the vibrationally coupled donor-acceptor model was proposed to give rise to significant memory effects that cannot be captured within a purely Markovian description. Here, we test this hypothesis directly by applying the NM-EFLD approach to the nonequilibrium transport regime and examining how the electronic memory influences the resulting vibrational dynamics.

In Fig.~\figref{fig: pos Delta NESS results}, we plot the nonequilibrium steady-state vibrational excitation $\langle N_{\text{vib}}^{\text{ss}}\rangle$ as a function of bias voltage and for $\Delta = 100\text{meV}$. Importantly, we show not only $\langle N_{\text{vib}}^{\text{ss}}\rangle$ obtained via the Markovian and non-Markovian EFLD approaches, but also from fully quantum HEOM simulations, which provides a numerically exact benchmark for this model. Details of the HEOM approach are given in Sec.~IV of the Supplementary Material. Also shown for comparison is the steady-state vibrational excitation for the scenario with no electronic energy splitting, $\Delta = 0$. As was discussed in Ref.~\cite{Preston2026_negative}, as well as previous investigations of similar models \cite{Lue2011,simine_vibrational_2012,segal_heat_2005,segal_heat_2006,segal_nonequilibrium_2011,Foti2018_origin,foti_interface_2017}, the inclusion of a positively-biased energy gap drives vibrational excitation beyond the standard heating that is observable in the $\Delta = 0$ case, as transporting electrons must inject on average $2\Delta$ of energy into the vibrational mode.

Although this nonequilibrium vibrational excitation manifests itself in a region of negative Markovian electronic friction, as shown in the red line of Fig.~\figref{fig: markovian electronic forces pos delta}, the Markovian electronic forces evidently do not capture the entire effect, as the M-EFLD approach actually underpredicts the quantum benchmark in the high-voltage limit. In contrast, the NM-EFLD approach yields steady-state vibrational excitation much closer to the quantum benchmark for $\Phi > 0.2\text{V}$. At this voltage, the unperturbed electronic sites enter the bias window and the mechanism of current-induced heating switches from being dominated by off-resonant stochastic heating to the deterministic effect described above. 

This result supports the analysis of Ref.~\cite{Preston2026_negative}, which argued that, since $\tilde{\gamma}(x,\omega)$ and $\tilde{D}(x,\omega)$ showed significant finite-frequency structure in nonequilibrium for the vibrationally coupled donor-acceptor model, non-Markovian effects in the electronic forces are likely to be important to the stationary dynamics. Indeed, Fig.~\figref{fig: electronic force spectrum x = 0} shows that while the Markovian electronic friction at $\Phi = 0.4\text{V}$ is negative, thus driving the vibrational mode, the finite-frequency contributions are negative and larger in magnitude, hence the increased vibrational excitation when these effects are included in the dynamics via the auxiliary variables. 

The importance of non-Markovian effects in the nonequilibrium steady-state observables is further illustrated in Fig.~\figref{fig: neg Delta NESS results (a)}, which shows the vibrational excitation as a function of bias voltage for $\Delta=-100\,\mathrm{meV}$ and $\Delta=0\,\mathrm{meV}$. Here, the negative electronic energy splitting suppresses vibrational excitation in comparison to the $\Delta = 0$ case, since transporting electrons must extract an energy of $2\Delta$ from the vibrational mode.

\begin{figure}
    \centering
    \includegraphics[width=\linewidth]{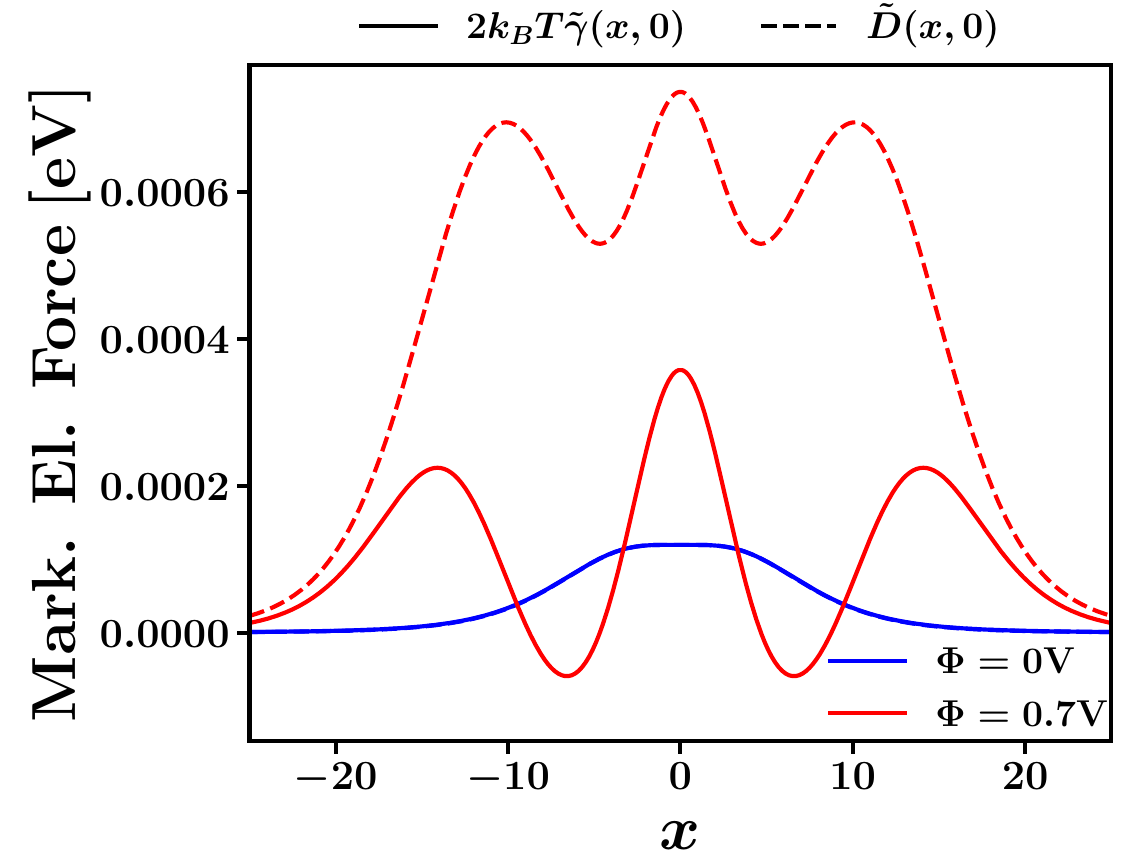}
    \caption{Markovian electronic friction (solid lines) and correlation function of the stochastic force (dashed lines) plotted as a function of vibrational coordinate, $x$, and for two voltages: $\Phi = 0\text{V}$ (blue) and $\Phi = 0.7\text{V}$ (red). The electronic energies are chosen as $\Delta = -100\text{meV}$.}
    \label{fig: markovian electronic forces neg delta}
\end{figure}

At low bias, shown in Fig.~\figref{fig: neg Delta NESS results (a)}, both the Markovian and non-Markovian EFLD approaches capture this effect well, as they closely match the numerically exact HEOM benchmark. Beyond a critical bias, however, $\Phi\approx0.6\,\mathrm{V}$, the two methods diverge dramatically, which is shown in Fig.~\figref{fig: neg Delta NESS results (b)}. The Markovian EFLD approach becomes unstable, predicting a nonequilibrium steady-state vibrational excitation orders of magnitude larger than the exact quantum result. In contrast, the NM-EFLD approach remains stable over the full bias range and continues to reproduce the quantum benchmark with excellent accuracy.

\begin{figure}
    \centering
    \phantomsubfloat{\label{fig: electronic force spectrum neg delta x = 0 (a)}}
    \phantomsubfloat{\label{fig: electronic force spectrum neg delta x = 0 (b)}}
    \includegraphics[width=\linewidth]{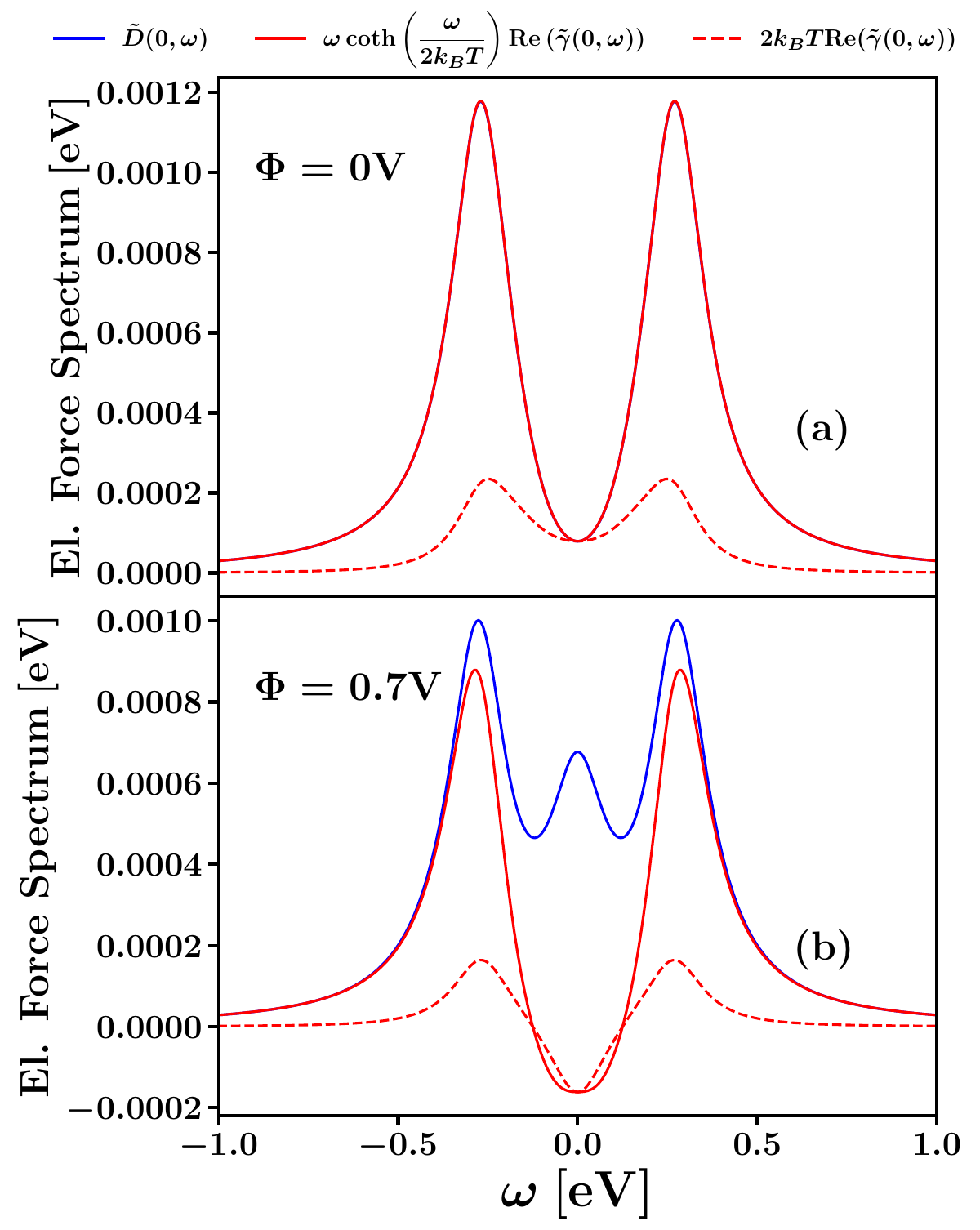}
    \caption{Exact frequency spectrum of the non-Markovian electronic forces at vibrational coordinate $x = 6.3$ and for voltages (a) $\Phi = 0\text{V}$ and (b) $\Phi = 0.7\text{V}$. The electronic energies are chosen as $\Delta = -100\text{ meV}$.}
    \label{fig: electronic force spectrum neg delta x = 6.3}
\end{figure}

To understand the origin of this instability, Fig.~\figref{fig: markovian electronic forces neg delta} shows the Markovian electronic friction and stochastic force correlation strength as functions of the vibrational coordinate for $\Delta=-100\,\mathrm{meV}$. At equilibrium ($\Phi=0\,\mathrm{V}$), the classical fluctuation-dissipation theorem is satisfied across the entire coordinate range. Under finite bias ($\Phi=0.7\,\mathrm{V}$), however, the stochastic force exceeds the electronic friction over a substantial region, $\tilde{D}(x,0) \geq 2k_{B}T\tilde{\gamma}(x,0)$, reflecting current-induced vibrational heating.

In contrast to the $\Delta>0$ case, the electronic friction is now positive around $x=0$, consistent with the physical expectation that electron transport damps the vibrational motion in this region. However, pronounced regions of negative Markovian friction emerge around $x\approx\pm6.3$, where the Markovian dynamics predicts that the electronic current injects energy into the vibration. These regions ultimately drive the instability observed in the Markovian simulations.

The origin of this apparent contradiction becomes clear upon examining the full electronic force spectrum. Fig.~\figref{fig: electronic force spectrum neg delta x = 6.3} shows that, although the zero-frequency (Markovian) friction is negative at $x=6.3$, the finite-frequency components, $\tilde{\gamma}(x,\omega\neq0)$, are positive and larger in magnitude. Consequently, the sign of the Markovian friction coefficient alone does not determine whether the electronic environment damps or drives the vibrational motion. Instead, the net energy exchanged over a vibrational cycle is governed by the full frequency-dependent friction, as demonstrated in Eq.\eqref{eq: dissipated power} and Eq.\eqref{eq: power density}.

These results demonstrate that, for the vibrationally coupled donor-acceptor model considered here, non-Markovian electronic forces are essential for accurately describing the nonequilibrium vibrational dynamics, even in the steady state. In doing so, they confirm the hypothesis of Ref.~\cite{Preston2026_negative} that the electronic relaxation is governed not only by the molecule-metal coupling, $\Gamma$, but also by the electronic energy splitting, $\Delta$. Because charge transport through the system occurs exclusively via the electronic-vibrational coupling, the electronic relaxation dynamics become intrinsically coupled to the vibrational motion. The resulting electronic back-action therefore retains memory of the vibrational trajectory over multiple characteristic timescales, rendering the Markovian approximation insufficient in regimes where these timescales become comparable. More broadly, these results demonstrate that retaining the full temporal structure of the electronic forces is essential whenever the electronic environment cannot be characterized by a single, rapidly decaying relaxation timescale.

\section{Conclusion} \label{sec: Conclusion}

Although electronic friction and Langevin dynamics provides a powerful framework for simulating the nonadiabatic vibronic dynamics of molecules at metal surfaces, it is most commonly applied in the Markovian limit. While this approximation is often adequate for systems with heavy nuclei or strong molecule-metal coupling, recent work has shown that many-body effects and bias-driven inelastic electronic transitions can significantly alter the electronic relaxation dynamics, compromising both the accuracy and stability of the Markovian EFLD approach, particularly in nonequilibrium and transient regimes. Propagating the corresponding non-Markovian Langevin equation, however, is challenging because the electronic friction and stochastic force exhibit coupled coordinate- and time-dependent memory, even within the quasi-stationary approximation.

To address this problem, we proposed a numerical framework for solving the non-Markovian Langevin equation based on Markovian embedding. By representing the electronic friction and stochastic force as independent sums of exponential functions, their non-Markovian influence can be captured through deterministic and stochastic auxiliary variables propagated alongside the physical vibrational degrees of freedom. In contrast to existing embedding approaches, our method treats both equilibrium and nonequilibrium quantum electronic noise while retaining the full coordinate dependence of the electronic forces. We refer to this approach as non-Markovian electronic friction and Langevin dynamics (NM-EFLD).

We presented the complete implementation of the method, including the exponential decomposition of the electronic friction and stochastic force, together with an efficient propagation algorithm based on an extended ABOBA integrator with an additional operator splitting for the auxiliary variables. Applying NM-EFLD to a vibrationally coupled donor-acceptor model of a molecular junction, we demonstrated that the method accurately reproduces numerically exact HEOM benchmark results in both current-induced vibrational heating and cooling regimes. In particular, NM-EFLD remained stable and quantitatively accurate in parameter regimes where the conventional Markovian EFLD approach became unstable, demonstrating that non-Markovian electronic forces play a decisive role in the nonequilibrium dynamics of this model.

More broadly, the proposed framework is not restricted to the donor-acceptor model considered here, but is applicable to a wide range of molecular systems interacting with metallic environments. Since it requires only the electronic friction kernel and the correlation function of the stochastic force as input, the method is compatible with a variety of electronic-structure and quantum transport approaches. Furthermore, although not explored in this work, the NM-EFLD offers direct axis to noise statistics in electronic observables in nonequilibrium, such as charge current, where non-Markovian effects are well known to play an important role. We therefore anticipate that NM-EFLD will provide a practical route to incorporating non-Markovian electronic effects into large-scale simulations of current-driven molecular transport, vibrational energy transfer, and surface scattering, where memory effects are expected to become increasingly important. 

\section*{Acknowledgements}
This work was supported by the Deutsche Forschungsgemeinschaft (DFG) within the framework of the Research Unit FOR5099 “Reducing complexity of nonequilibrium systems”. R.J.P thanks the Alexander von Humboldt Foundation for the award of a Research Fellowship. The authors acknowledge the support by the state of Baden-Württemberg through bwHPC and the DFG through Grant No. INST 40/575-1 FUGG (JUSTUS 2 cluster).

\clearpage

\bibliography{Main_text_incl._fig.bib} 

\end{document}